%% file: _FAU-CS-TR_deus_tex.tex
\documentclass[11pt,a4paper,conference,twoside]{IEEEtran}
\pdfoutput=1 % pdflatex hint for arxiv.org (within first 5 lines)
\usepackage{cs-tr}

\usepackage{ifpdf}

\ifpdf
\else
  \usepackage[inactive, nowinedt]{srcltx}
\fi

\ifpdf
	\input glyphtounicode
\fi

\usepackage{cite}

\usepackage{acronym}

\usepackage{graphicx}

\usepackage[tight,hang,raggedright]{subfigure}
\usepackage{url}

\usepackage{listings,color}
\definecolor{darkgrey}{rgb}{0.95,0.95,0.95}
\acrodef{PIM}{Platform Independent Model}
\acrodef{RHIN}{Regional Healthcare Information Networks}
\acrodef{EHR}{Electronic Health Record}
\acrodef{NHS}{National Health Service}

\usepackage{xspace}
\newcommand{\hlseven}{\textsc{hl7}\xspace}
\newcommand{\ihe}{\textsc{ihe}\xspace}
\newcommand{\sciphox}{\textsc{sciphox}\xspace}
\newcommand{\omg}{\textsc{omg}\xspace}
\newcommand{\rhin}{RHIN}
\newcommand{\hygeianet}{\textsc{hygeia}net\xspace}

\newcommand{\ehr}{\textsc{ehr}\xspace}
\newcommand{\phr}{\textsc{phr}\xspace}

\newcommand{\pdf}{\textsf{PDF}\xspace}
\newcommand{\xml}{\textsf{XML}\xspace}
\newcommand{\iiop}{\textsf{IIOP}\xspace}
\newcommand{\ejb}{\textsf{EJB}\xspace}
\newcommand{\rmi}{\textsf{RMI}\xspace}
\newcommand{\soap}{\textsf{SOAP}\xspace}
\newcommand{\wsdl}{\textsf{WSDL}\xspace}
\newcommand{\wsstar}{\textsf{WS-*}\xspace}

\newcommand{\xmpp}{\textsf{XMPP}\xspace}
\newcommand{\dotnet}{\textsf{.Net}\xspace}
\newcommand{\ebxml}{\textsf{ebXML}\xspace}
\newcommand{\spring}{\textsf{Spring}\xspace}
\newcommand{\osgi}{\textsf{OSGi}\xspace}

\newcommand{\rest}{\textsc{rest}\xspace}
\newcommand{\restful}{\textsc{rest}ful\xspace}
\newcommand{\dmps}{\textsc{dmps}\xspace}
\newcommand{\deus}{\textsc{deus}\xspace}

\newcommand{\uml}{\textsc{uml}\xspace}

\newcommand{\pids}{\textsc{pids}\xspace}
\newcommand{\cda}{\textsc{cda}\xspace}
\newcommand{\rim}{\textsc{rim}\xspace}
\newcommand{\xds}{\textsc{xds}\xspace}
\newcommand{\atna}{\textsc{atna}\xspace}
\newcommand{\pix}{\textsc{pix}\xspace}
\newcommand{\dicom}{\textsc{dicom}\xspace}

\newcommand{\icd}{\textsc{icd}\xspace}
\newcommand{\snomed}{\textsc{snomed}\xspace}

\newcommand{\loinc}{\textsc{loinc}\xspace}

\newcommand{\dc}{digital card\xspace}
\newcommand{\dcs}{digital cards\xspace}
\newcommand{\pif}{personal information file\xspace}
\newcommand{\dif}{distributed information folder\xspace}
\newcommand{\fif}{foreign information file\xspace}

\newcommand{\hcis}{HCIS\xspace}

\usepackage{color}
\trauthor{Christoph P. Neumann, Florian Rampp, Richard Lenz}
\tremail{christoph.neumann@cs.fau.de}
\trgroup{Institute of Computer Science 6 (Data Management)}
\trtitle{DEUS: Distributed Electronic Patient File Update System}
\trreport{CS-2012-02}
\trmonth{March}
\tryear{2012}

\begin{document}

\trmaketitle

\author{\IEEEauthorblockN{\thetrauthor}%
\IEEEauthorblockA{%
 	\thetrgroup \\
 	Dept. of Computer Science, University of Erlangen, Germany \\
 	\texttt{\thetremail}
}}
%\title{\thetrtitle}
\title{DEUS: Distributed Electronic Patient File\\Update System}

\maketitle

% the text begins here
%%%%%%%%%%%%%%%%%%%%%%% /TR %%%%%%%%%%%%%%%%%%%%%%%%%%
\input{00_abstract}

%%%%%%%%%%%%%%%%%%%%%%% INCLUDES %%%%%%%%%%%%%%%%%%%%%%%%%%
\input{10_prearch.tex}
\input{50_architecture.tex}
\input{60_postarch.tex}
%%%%%%%%%%%%%%%%%%%%%%% /INCLUDES %%%%%%%%%%%%%%%%%%%%%%%%%%

%ACKNOWLEDGMENTS are optional
%\iffalse
%\section{Acknowledgments}
%The first author wants to thank Dr.~med.\ Helmut Neumann who as a gynecologist explained
%breast cancer treatment to me and Rita M.\ Neumann who survived breast cancer
%and familiarized me with the patient perspective.
%\fi

%\vfill
%\pagebreak

\bibliographystyle{IEEEtran} % plain, abbrv, alpha, unsrt
\bibliography{promed_global,cpn_self_all,cpn_studis_all,cpn_supervised_all,cpn_coops_all}  % sigproc.bib is the name of the Bibliography in this case
% ACM needs 'a single self-contained file'!
%
%%%APPENDICES are optional
%%\balancecolumns
%\vfill{}
%\pagebreak
%\appendix
%\input{90_appendix.tex}

\end{document}

%% file: 00_abstract.tex
%auto-ignore
\begin{abstract}

Inadequate availability of patient information is a major cause for medical errors and affects costs in healthcare.
Traditional approaches to information integration in healthcare do not solve the problem.
Applying a document-oriented paradigm to systems integration enables inter-institutional information exchange in healthcare.
The goal of the proposed architecture is to provide information exchange between strict autonomous healthcare institutions, bridging the gap between primary and secondary care.

In a long-term healthcare data distribution scenario, the patient has to maintain sovereignty over any personal health information.
Thus, the traditional publish-subscribe architecture is extended by a phase of human mediation within the data flow.
\deus{} essentially decouples the roles of information author and  information publisher into distinct actors, resulting in a triangular data flow.
The interaction scenario will be motivated.
The significance of human mediation will be discussed.
\deus{} provides a carefully distinguished actor and role model for mediated pub-sub.
The data flow between the participants is factored into distinct phases of information interchange.
The artefact model is decomposed into role-dependent constituent parts.
Both a domain specific (healthcare) terminology and a generic terminology is provided.
From a technical perspective, the system design is presented.
%%
%It is composed into subsystem modules that match the prior conceptual decomposition into roles, phases, and artefacts.
%%
The sublayer for network transfer will be highlighted as well as the subsystem for human-machine interaction.
\end{abstract}

\begin{keywords}
Healthcare, information systems, inter-institutional, domain engineering, distributed applications, distributed data structures, doc\-u\-ment-orientation, publish-subscribe, human mediation
\end{keywords}

%% file: 10_prearch.tex
%auto-ignore
\section{Introduction}

In a systems analysis of adverse drug events, 18\% of the medical errors were associated with inadequate availability of patient information \cite{leape1995saa}.
The problem of inadequate availability of patient information %, such as prescriptions or the results of laboratory tests, 
as a major cause for medical errors is aggravated by the rise of healthcare networks and the increasing number of healthcare parties that are involved in a treatment:
The aging of western society affects the public health sector, chronic diseases and multimorbidity become the focus of interest, and the cost pressure rises.
Chronic diseases and multimorbidity, like cancer, diabetes, asthma, and cardiac insufficiency, require more healthcare parties than common diseases.
Coevally, the rapid advance in medicine leads to an advancing specialization of physicians that is an additional cause for the increasing number of involved parties regarding a single patient's treatment.
For improving the treatment quality and in order to avoid unnecessary costs, an effective information and communication technology is vital for the support of inter-institutional patient treatment.

An IT infrastructure for healthcare networks must respect and consider the autonomy of preexisting systems in different institutions.
At the same time some kind of integration is required, which helps to preserve inter-institutional data consistency.
Closely integrated systems with a common database and terminological standards for database contents are unrealistic in this scenario.
%%
%Instead, a platform is required, which supports semantic agreements among the communicating parties.
%%
%Ideally (but not necessarily), such an agreement could be based on standard domain ontologies, which contribute to semantic compatibility.
%%
In particular, strict autonomy of the institutions requires the abdication of central infrastructure like joint databases, transaction monitors, and central context managers.
Shared communication requires minimal standards avoiding full-fledged platform-specific middleware frameworks.
Instead, a document-oriented publish-subscribe paradigm %\cite{Beyer2005}
is favored, which supports loose coupling of systems at different sites.
%
%Document-ori\-en\-ta\-tion favors local autonomy over central hegemony. % adhering to the design goal of loose coupling.

In any case, semantic agreements like healthcare ontologies, terminologies, and clinical models evolve over time.
Therefore, an adequate system design methodology for evolutionary information systems becomes imperative \cite{Lenz2003a}.
A layered approach for healtcare information sytems decouples system design processes on different levels of abstraction, decreases complexity in each layer, and thereby supports system evolution and responsive infrastructures. In  \cite{Lenz2006} such an approach with four layers is proposed: 
\emph{generic framework} layer, \emph{domain framework} layer, \emph{application layer}, and \emph{custom layer}, with different people as key drivers for the different layers. 
This model, however is an idealized picture, which is not yet related to existing standards and frameworks. 
In this paper, the research focus is on domain framework solutions as extensions to existing healthcare information systems.

This paper describes a publish-subscribe architecture for healthcare supply chain scenarios. % for inter-institutional information exchange in healthcare.
This technical report is an extended version of a previous publication \cite{NRDL09deus}.
There are two distinguishable features of the proposed solution:
to apply document-orientation as instrument of inter-institutional integration and to allow patients to control information distribution.
To put a mediated publish-subscribe architecture into practice requires a systematic distinction of actors, roles, phases, and responsibilities in the distribution scenario.
%%
%%
%The benefits of document-orientation will be discussed in section \ref{section-document-orientation}.
%%
%The relation between the exchanged artifacts and the actors as well as their responsibilities is detailed in section \ref{section-proposed-solution}.
%%
%The realized system architecture with its modularization and its plug-in mechanisms for arbitrary transfer protocols is outlined in section \ref{section-deus-architecture}.
%%
%The primary goal is to foster inter-institutional information exchange in healthcare, bridging the gap between institutions of the primary and secondary care; in section \ref{section-defense} the solution is defended in contrast to centralized integration solutions, and section \ref{section-related-work} discusses related work in the field of distributed approaches.
%%
%%
The proposed architecture essentially decouples the roles of information author and information publisher into distinct actors.

\section{Supply Chains in Healthcare}
\label{section-Healthcare}

A short overview of the domain participants is given:
The focus of the medical supply chain in Germany are the patients who are treated by office-based physicians foremost, collectively described as the \emph{primary care}.
The \emph{secondary care} adds hospitals, laboratories, pharmacies, and ancillary medical institutions as participants of the medical supply chain. 
Accompanying participants are the health insurance funds and the associations of statutory health insurance physicians.

The whole interaction and collaboration is liable to many technical, organizational, economic, and legal factors.
The legal boundary conditions are critical for the information provision and availability because warranty of data protection is essential for patient-related data.
The patient has to maintain sovereignty over any personal health information.
This provides a basic motivation for the mediation approach that will be applied to the publish-subscribe pattern.

\subsection{Inter-Institutional Problems}

Considering the comprehensive medical supply chain, functional integration and process integration between the autonomous information systems of the several participants is still unsolved in organizational and diagnostic-therapeutic processes \cite{WTB+2008}.
Particularly chronic diseases like diabetes, asthma, and cardiac insufficiency require a long-cyclic exchange of patient information between all involved healthcare professionals from different institutions \cite{LSLG200l}.
At present, this is done by paper documents like discharge letters or by repeated anamnesis interviews with the patient.
The sticking point is that the discharge letters are frequently missing or are insufficient in detail \cite{TBBT2002}, either because they are not written by the physicians at all or because they are not available to all involved parties \cite{Same2004}.
Repeated anamnesis interviews are no alternative for document interchange between healthcare professionals.
Redundantly applied diagnostic methods by each distinct institution are the norm.
As simple as order entry and result reporting may seem, it is still one of the most important issues in healthcare information systems \cite{LBM+2005}.
The data integration of healthcare information systems will be addressed by a document-oriented approach in the proposed solution.

In order to foster the continuity of care \cite{FMF+84}, the inter-insti\-tu\-tional cooperation needs to bridge the current gap between institutions of the primary and secondary care \cite{MUBP2005}.
Such effort must not instrument regional standards, as it is done in \emph{regional healthcare information networks} (RHIN) \cite{WVVM2001}, but \emph{transregional standards}.
Accomplishing information exchange in distributed healthcare scenarios requires the integration of heterogeneous and \emph{strict autonomous IT systems}.
To allow for \emph{inter-institutional cooperation} the support of distributed and seamless flow of information is required, thus changing paradigms from closed and hegemonic to open and distributed architectures.
The proposed architecture adheres to these boundary conditions.

\subsection{From Bilateral Information Exchange\\to Information Distribution}

Information interchange by letters is the way of traditional cooperation -- a referral from one institution to another delegates responsibility and liability of diagnosis or therapy to the other institution.
Yet, genuine physician teams from different institutions are upcoming \cite{RoLa2003}.
For some years now, in Germany, the treatment of breast cancer is organized by accredited in-station breast cancer treatment centers cooperating with manifold accredited partners like oncologists, radiologists, and the post-operative care \cite{BKB+2003}.
Collaborative treatment scenarios can be described as distributed medical treatment processes with physician teams from different institutions interacting closely meshed for treating complex chronic diseases and multimorbidity.

Such scenarios change the requirements for the availability of patient information and deserve extended information exchange models, still adhering to the strict autonomy of the involved IT systems.
Although comprehensive IT support for such closely meshed treatment scenarios in general has unsolved legal boundary conditions, this paper outlines an architecture for distributed patient files as a technical foundation.
It is based on digital information cards that are yielded from institutional \emph{electronic health records} (\ehr{}), e.g. \cite{PoBu2005}, into a distributed publish-subscribe system that allows the patient to govern data interchange.

%%% FORCING THE PICTURE AT THE SIDE OF SUBSECTION ``Phases of Information Interchange''
\begin{figure*}[ht] \centering 
\includegraphics[width=.85\textwidth]{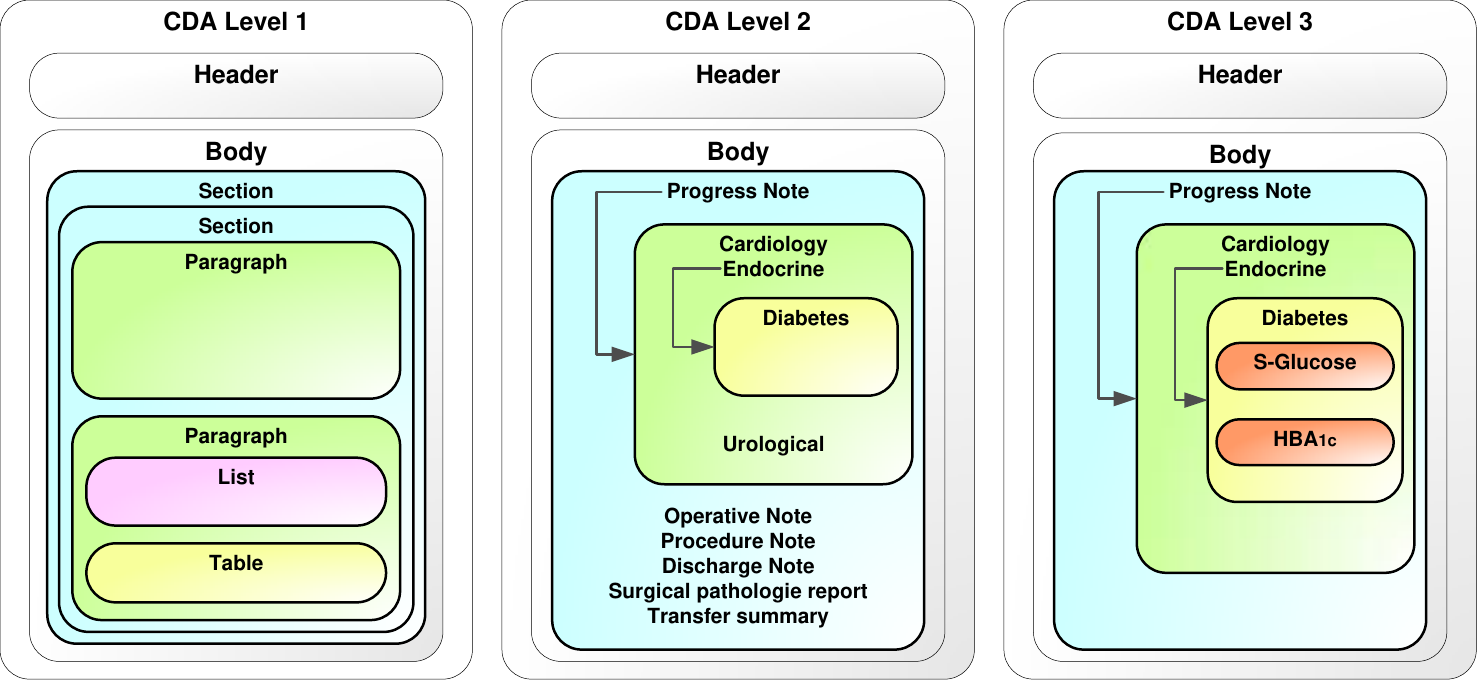}
    \caption{HL7 v3 CDA structure outline for levels 1, 2, and 3 (cf.\,\cite{schuler2002tutorial,sippel2005da})}
    \label{figure-CDA-levels}
\end{figure*}

\subsection{Fundamentals of Healthcare Data Models}

Exchange of patient information among institutions requires data compatibility.
\emph{Data integration} is the process that achieves such data compatibility, either by common standards or by data transformation.
Data integration for medical processes requires standards for medical terminology that have to deal with volatile medical concepts \cite{MOD+1998}.
Over the intervening years numerous standards for medical ontologies have been created on \emph{type level} for system implementers at design-time and on \emph{instance level} primarily for end-users at run-time.

At instance level, standards like \icd{} \cite{Zais1996}, \snomed{} \cite{SPSW2001}, and \loinc{} \cite{FMD+1996} exist which unremittingly evolve over time.
The \hlseven{}\footnote{Health Level 7, \url{http://www.hl7.org}} v2 \cite{HL7v26} is a well-established standard for clinical message specification.
It is a standard on type level, and incorporates arbitrary coding schemes and terminologies on instance level.
Dealing with inherent volatility of reference terminologies by information system design in a general matter actually is another unsolved scientific issue.
Despite many attempts, there is no stable unique and comprehensive ontology of the medical domain in sight.
%
%Nonetheless, effective systems can be achieved with available standards \cite{BANT2006}.

The \hlseven{} v2 standard allows for the specification of arbitrary self-defined messages, which has lead to incompatibilities.
%
%A \emph{Canonical Data Model}, as it is described as enterprise integration pattern in \cite{HoWo2003}, ...
%
The relatively new \hlseven{} v3 standard is based on the \hlseven{} v3 \emph{reference information model} (\rim{}) and is radically different from the v2 standard:
It allows for new types to be derived from a limited number of core classes, enabling \rim{}-based systems to handle even unknown message-types in a generic way.
Furthermore, a conceptual change from messages to documents is provided by the \hlseven{} v3 \emph{clinical document architecture} (\cda{}) substandard \cite{DAB+2001}.
\cda{} allows for \xml{}-structured medical documents.
Based on \hlseven{} v3 \cda{}, in Germany, the \sciphox{}\footnote{Standardized Communication of Information Systems in Physician Offices and Hospitals using \xml{}} \cite{HeSD2003} 
working group tries to develop specific document content types for German healthcare; for example, referral vouchers and discharge letters\footnote{Particularly ``eArztbrief SCIPHOX CDA R1'' and its advancement ``eArztbrief VHitG CDA R2'' \cite{eArztbrief}}.

Any new standards should respect the ones already in practice for backwards-compatibility and to achieve and maximize acceptance.
In conclusion, the event type of the proposed mediated publish-subscribe system will be based on \hlseven{} v3 \cda{}.

\subsection{Objectives}

There are two prime objectives of the proposed solution.
The first is the abdication of any central server. % as adherence to the strict autonomy of the institutions.
The second is the application of document-oriented integration with lightweight interfaces instead of service-oriented integration with semantically rich interfaces.
A subsequent design objective is to aim for minimal standards in order to yield minimal requirements to the participating systems. %as well as coarse-grained and therefore intuitional models for the organizational and medical process support, adhering to the design goal of loose coupling.
Favoring local autonomy over central hegemony requires, for example, that distribution of information will not be enforced, but is voluntary and process participation can be supplemented on demand.
Platform independence and the avoidance of vendor lock-ins require that the basic architecture is decoupled from any specifically instrumented middleware and components off-the-shelf.
A sophisticated system modularization has been a major design goal.

\section{Applying Documents as Event Type}
\label{section-document-orientation}

Integration in healthcare is traditionally based on in\-ter\-face-orientation.
Three-tier network-based architectures with remote procedure calls are yet the dominant style of information systems.
The most common technological occurrence of remote invocations is based on \soap{} with \wsdl{}, augmented by several \wsstar{} frameworks.
The \emph{interface-oriented integration} focuses on available functionality, and the integration method affects semantically rich interfaces.
An invocation uses parameters to detail its synchronous service request to a target system.
In interface-oriented integration the information being passed is not necessarily viable on its own but often in the context of the service request only.

Even if a service is triggered event-oriented using asynchronous messaging, like it is done in \hlseven{} v2-based systems, such parameters or messages essentially represent transient fine-grained information that is assimilated by the targeted system.
The three main problems in information integration projects, including healthcare systems, are insufficient synchronization of redundant data, problems with data consistency, and functional overlapping \cite{LeKu2003}. %, and inadequate process support.
Therefore, interface-oriented and message-oriented integration between distinct institutions is complex and custom-designed.

In contrast, documents are coarse-grained, self-contained, and viable.
A document carries its own context information and can exist independently from the system it stems from.
Changes are not propagated by update information, but by creating an updated document that replaces its predecessor.
The \emph{document-oriented integration} focuses on available information, and the integration method affects the semantic scalability of document models, using standardized and minimal interfaces for hand-over.
Redundancy in data distribution is not critical with documents because, due to the self-con\-tain\-ed\-ness, a synchronization in the classical sense is not required.
Likewise are data consistency checks confined to the scope of the document.

\hlseven{} v3 \cda{} provides semantic scalability for healthcare documents, both because this has been an inherent feature of the underlying \rim{} and because \cda{} is particularly structured in three levels of semantic abstraction:
In Fig.\,\ref{figure-CDA-levels} a basic outline of the three \cda{} levels is provided as illustration.
\cda{} level 1 is the unconstrained \cda{} specification.
\cda{} level 2 applies section-level templates.
\cda{} level 3 applies entry-level templates.
For example, \cda{} level 1 simply ensures the ability to display a document like a \pdf{} file.
Any \cda{} document can be accepted without immediate support for processing.
Advanced semantic processing support of \cda{} level 2 or 3 can be added to the system, seamlessly enhancing the information value of already stored \cda{} documents.

The \emph{deferred system design} principle of evolutionary systems \cite{Pate2002} requires semantic decisions not to be frozen in an interface schema because they are hard to revise.
\hlseven{} v3 \cda{} supports deferred system design by its semantic scalability.
The proposed solution uses \cda{} for typing the \emph{Document Messages} \cite[p.\,147]{HoWo2003} in its mediated publish-subscribe scenario.
Event models have a major impact on the flexibility and usability \cite{RoSS2007}; applying a document-oriented approach improves the adaptability of the systems by deferring schema decisions from design-time to deploy- or run-time \cite{Lenz2009yb}.

\section{Proposed Solution}
\label{section-proposed-solution}

The proposed solution architecture, the \emph{distributed electronic patient file update system} (\deus{}), 
applies the doc\-u\-ment-oriented idea in form of \deus{} \emph{digital cards}\footnote{Unfortunately, the preferred term \emph{information card} has been patented in 1972 by Paul P.\ Castrucci, US patent 3702464.}: \dc{}s are self-con\-tained and viable containers of information.
A \dc{} is authored by an information provider (first part of the \dc ID; e.g.~a physician) and it concerns a person (second part of the \dc ID; e.g.~a patient).
Because an author can provide several \dc{}s about a concerned person, the author provides an additional discriminator as third part of its identifier.
The \dc{} as information container is subject to PKI signatures, as it will be detailed by the \deus{} scenario description.
\deus{} supports even closely meshed physician teams and fosters trans-sectional, life-long, and patient-centered healthcare documentation.

From a conceptual perspective, such exchanged information is a document as required by a doc\-u\-ment-oriented architecture approach.
However, the intended granularity of a \deus{} \dc{} is a more fine-grained one than the one that is experienced from paper-based working practice in healthcare.
This improves the structure of the patient files, and provides higher selectivity in retrieval and display.
One \deus{} \dc{} contains, for example, a diagnostic finding, clinical evidence, a diagnosis, a therapeutic measure, an order, or a prescription.
The \dc{} metaphor has been inspired by the Higgins project \cite{HigginsProject} with its I-Cards as foundation of an open source identity framework.
The exemplary actor ``Dr.\,Higgins'', as actual contributor of \dc{}s in the \deus{} scenario in the next section, is a homage to this project.

\subsection{Interaction Scenario}

\begin{figure}[ht] \centering 
\includegraphics[width=1\columnwidth]{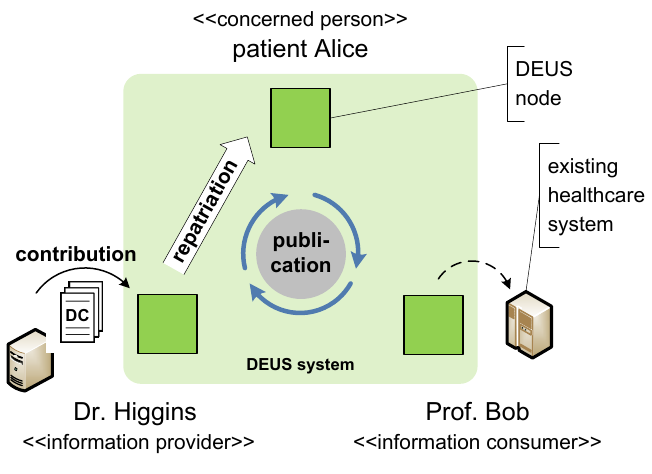}
    \caption{The DEUS scenario as mediated publish-subscribe system}
    \label{figure-DEUS-scenario}
\end{figure}

The basic \deus{} scenario is outlined in Fig.\,\ref{figure-DEUS-scenario}.
The patient (``Alice'') has recently visited a healthcare professional (``Dr.\,Higgins'') and the obtained information has to be shared to other involved parties (inter alia ``Prof.\,Bob'').
The local \emph{healthcare information system} (\hcis{}) of Dr.\,Higgins, the author of the obtained information, bundles the information into a \dc{}.
This \dc{} is electronically signed by its contributor and becomes the subject of information distribution.
Subsequently, it is contributed into the node's local \deus{} system extension.

The exported \dc{} is subsequently transfered to the account of patient Alice who is the person being concerned by the medical information.
The patient as sovereign of information distribution decides whether the information is accepted into the pool of \dc{}s that builds his or her \emph{personal patient file}.
The process of transfering a \dc{} from the contributing \deus{} system to the patient's \deus{} system together with the patient's decision about the acceptance of the \dc{} is named \emph{repatriation}.
Subsequently, the \dc{} is published to any subscriber \deus{} systems, like Prof.~Bob.
He will consume the information later, for example when Alice is visiting next time.

Each \deus{} participant owns a \deus{} account.
A \deus{} node is a healthcare information system with an installed \deus{} extension.
A \deus{} node can host multiple \deus{} accounts. %, and the applied data architecture implements multitenancy.
For the mediated publish-subscribe interactions, it is transparent whether an account resides on the same or another \deus{} node.

\subsection{The Significance of Human Mediation}

``Patient Empowerment'' is a general issue in eHealth, for example being addressed by the EU in form of a ministerial declaration \cite{euministererklaerung2003}.
Empowering the patient specifically to control information distribution might still seem radical:
Today, in working practice, the healthcare professionals can request patient information from each other.
The cornerstones are the thorough trust of a patient in his or her physician and the trust of physicians among each other, due to the Hippocratic oath.
Information exchange does not necessarily involve patients, aside from instrumenting the patient as a useful surrogate for postal delivery.

Life-long, trans-sectional, electronic storage and communication cannot be compared to such traditional approach:
If a physician requests information in form of a paper-based, signed request it is checked and kept filed by the addressed institution.
A plausibility check by a human actor is integral part of any such exchange, and any abuse in paper-based scenarios using faked requests risks instant detection; especially if more than a few patients are concerned.
%%
%Electronic approaches require basic education in IT security by the end-users, and are prone to overcredulous trust in IT-provided information.
%%
%Furthermore because the necessary IT education is not present
%%
%%Automating information publication of sensible patient-related information must not allow uncontrolled publication.
%
In contrast to paper-based requests, in electronic patient file scenarios, a high quantity of patients being affected by an attack will not automatically result in a timely detection, if the system-supported communication bandwidth and latency allows for fast assault automation.

%%% Forced above ``Actors and Roles''
\begin{figure*}[ht] \centering 
\includegraphics[width=.85\textwidth]{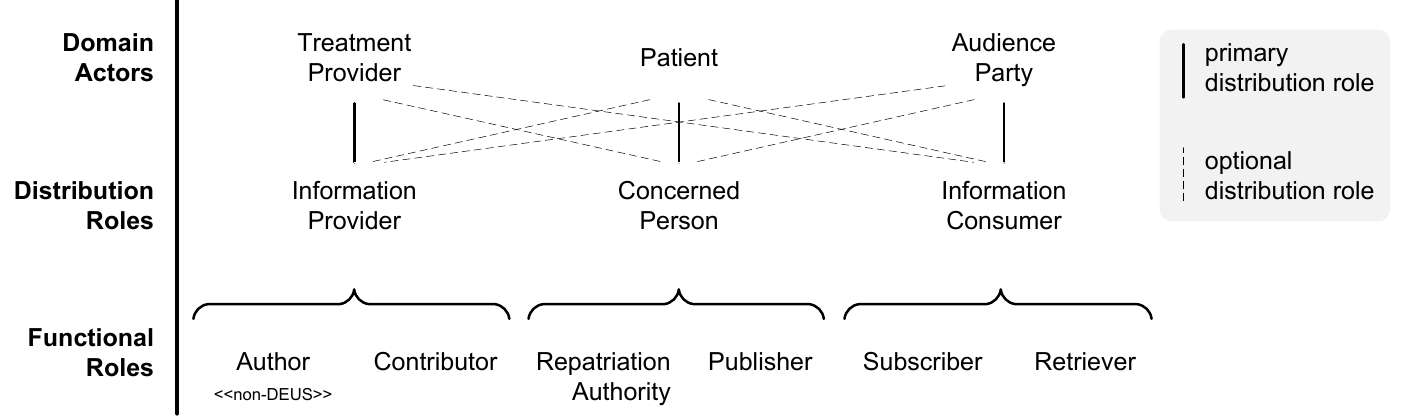}
    \caption{The DEUS roles of healthcare participants}
    \label{figure-deus-roles}
\end{figure*}

In conclusion, the life-long distribution of patient-related information requires a human authority over in\-ter-in\-sti\-tu\-tion-ally shared information, so that the % 
induced transparency of patient information by a server-side electronic storage does not erode doctor-patient confidentiality.
The only appropriate healthcare participant for publication control is the patient as the concerned person.
If the patient lacks the abilities to sovereign his or her healthcare information, it is possible to delegate this role to a legitimate proxy person or institution, possibly a general practitioner.
The mediation of \hcis{}-to-\hcis{} communication by the patients' \deus{} accounts reintroduces interception and control by a conscious human in an electronic environment for gradually automated information interchange.

\subsection{The Significance of Repatriation}

The necessity for a repatriation mechanism can further be clarified by describing its notable absence in a common distributed information environment -- the PGP key server infrastructure: 
It is possible for a rogue signer to attack a PGP key that is published to the global directory by creating thousands of garbage keys, signing the attacked key with all of them, and posting the attacked key with a bloated signature extension to the key server network. 
A PGP key server does not check whether a key is submitted by the owner of the key, but allows anyone to post additional key extensions to the key server network.
Henceforward, each key-updating owner and consumer is plagued by the bogus signatures, possibly leading to an unusable key and therefore successful denial of service.
An additional phase has not been considered in PGP key server infrastructure -- allowing only key extensions that are counter-signed by the owner for upload to the key server network would have solved the issue, implicating a kind of ``repatriation''.
Luckily, PGP users seldom attract attackers that would exploit this floodgate design flaw.

Distributed healthcare infrastructure for long-term information must preclude attackers from injecting bogus information by implementing a kind of repatriation.
This allows the concerned person to check a \dc{} received from an information contributor.
The concerned person will counter-sign an accepted \dc{}, because this enables the subscribers to verify the authenticity, finally.
A public key infrastructure that allows the concerned person to securely white list author accounts as valid information contributors optionally helps to automate the repatriation decision.
The proposed \emph{mediated publish-subscribe architecture} essentially decouples the roles of information author and information publisher into separate actors.
A description in subsect.\,\ref{sect-closed-comm-env} will portray how \deus{} could be used in organizationally closed environments with special trimmed-down logical patient accounts for scenarios in which patient interaction is not feasible.

\subsection{Actors and Roles}

The three basic participating actors and their different roles in the \deus{} information interchange are summarized in Fig.\,\ref{figure-deus-roles}.
Each actor has its dominant role in the distribution process, visualized as straight vertical lines between the layer of \emph{domain actors} and the layer of \emph{distribution roles}.
However, each actor can assume, by its \deus{} account, each distribution role as it is visualized by the clashed lines:
For example, the patient can act as information provider about himself, providing information like allergies or legacy paper documents that he contributes in transcribed or scanned form.
On the other hand, a treatment provider can use its \deus{} account in the role of a concerned person to provide and publish business card information or consultation hours information.
Even associations of statutory health insurance physicians could participate by contributing certificates like the ones required for accredited in-station breast cancer treatment centers, and health insurance funds could contribute the patient master data.

The distribution roles are related to the specific role that a \deus{} account assumes in an overall distribution scenario. %, interacting with other \deus{} accounts which reside on arbitrary nodes.
The \emph{functional roles} are responsibilities that are deduced from a distribution role.

%%% FORCING THE PICTURE AT THE SIDE OF SUBSECTION ``Phases of Information Interchange''
\begin{figure*}[tb] \centering 
\includegraphics[width=0.85\textwidth]{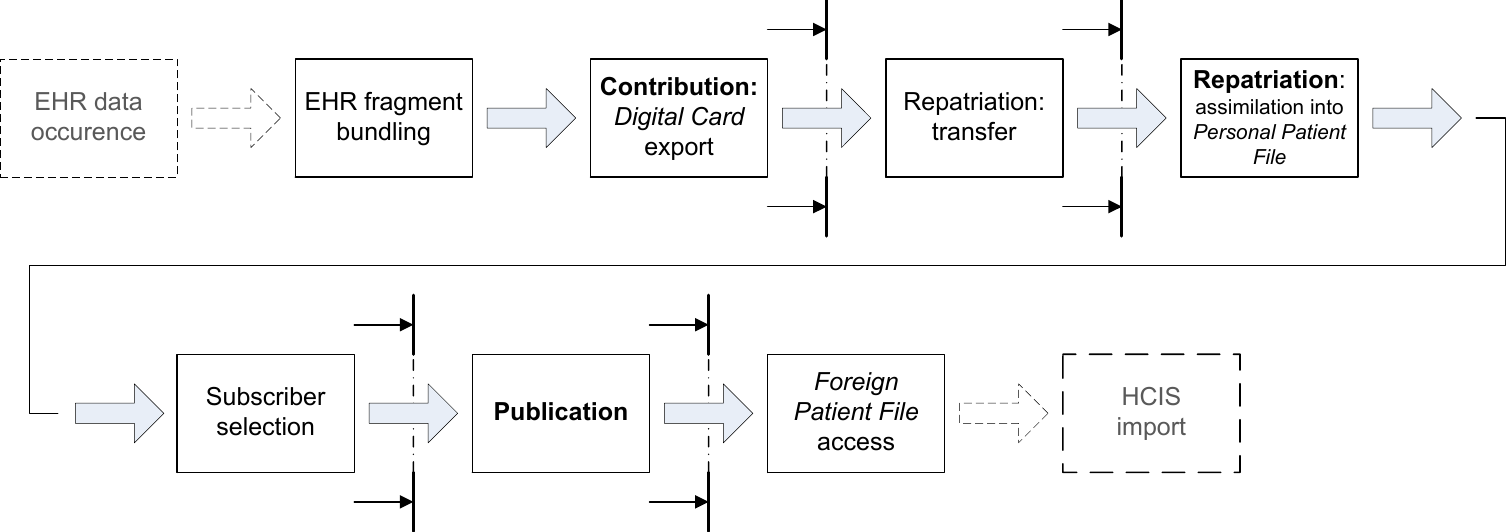}
    \caption{The DEUS contribution-repatriation-publication chain}
    \label{figure-deus-contributionchain}
\end{figure*}

The information provider acts as author of the information, which takes place inside the HCIS and is not part of \deus{}.
The information provider acts as contributor by exporting a \dc{} and handing it over to the actual \deus{} account.
The concerned person acts as repatriation authority by deciding about the validity of a repatriated \dc{}.
The concerned person acts as publisher by applying the selection of subscribers and performing the publication transfer.
The information consumer acts as subscriber by establishing subscriptions to the account of the concerned person and by accepting published \dc{}s.
Finally, the information consumer acts as retriever by accessing the information pool.
Each functional role will further be detailed by the equivalent subsystems in the \deus{} architecture description.

\subsection{Phases of Information Interchange}

The three basic phases in \deus{} information interchange are \emph{contribution}, \emph{repatriation}, and \emph{publication}.
Fig.\,\ref{figure-deus-contributionchain} describes an information chain from the perspective of a new patient-related information in an \ehr{} of a selected \hcis{}.
The vertical separators delimit system borders: 
The \ehr{} with the bundling component and export mechanism for \dc{}s resides in the contributor environment.
The repatriation transfer is done by point-to-point network infrastructure.

The repatriation acceptance and assimilation is handled in the environment of the concerned person that also comprises the subscriber selection:
A group-based selection is considered that allows to group physicians into teams so that repatriated \dc{}s are only published to team members.
Enabling subscribers to express predicates over classifying attributes or over the content of the \dc{}s, providing channel- or content-based subscriptions \cite{CaWo2002}, is not supported due to tremendous unsolved medical and legal implications.

%The publication transfer is provided by a pub/sub network infrastructure.
%%
%Because of the required platform independence, the \deus{} architecture is not coupled to a specific transfer protocol but allows for arbitrary pub/sub or event-bus integration.

The published \dc{} becomes part of the \emph{foreign patient file}, the read-only subset of the personal patient file at subscriber-side.
In distributed environments, a released information cannot be revoked completely: Any receiver has the possibility to subsequently process the received information in its local systems (last optional step of \hcis{} import in Fig.\,\ref{figure-deus-contributionchain}).

\subsection{Data Architecture}

\begin{figure*}[ht] \centering 
\includegraphics[width=.75\textwidth]{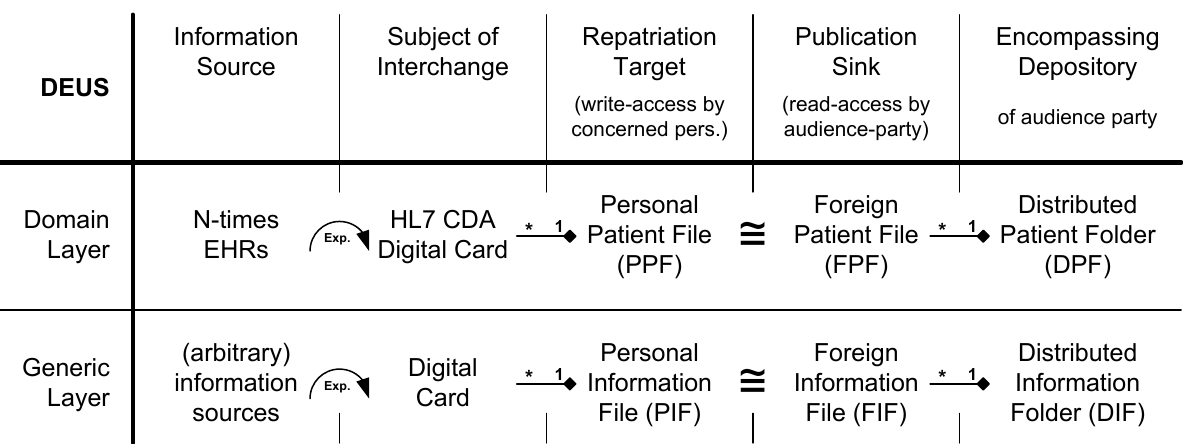}
    \caption{The DEUS artifacts}
    \label{figure-deus-artifacts}
\end{figure*}

The overview of the basic artifacts in \deus{} is provided in Fig.\,\ref{figure-deus-artifacts}.
The healthcare \emph{domain layer} artifacts are digital cards based on \hlseven{} \cda{}.
A \dc{} is the subject of interchange.
The \emph{personal patient file} is the central patient-oriented information hub.
The \emph{foreign patient file} is the read-only representation at subscriber-side, being a collection of digital cards.
If the patient manages distinct physician teams, the foreign patient files are a team-specific subset of the original comprehensive personal patient file.
The \emph{distributed patient folder} is managed by \deus{} and consists of all subscribed foreign patient files of a \deus{} account.

%%% FORCING THE PICTURE AT THE SIDE OF THE SUBSECTION ``DATA ARCHITECTURE''

The identity of a \deus{} account is essential to the addressing scheme of the artifacts because the \deus{} account IDs from different actors take part in several artifact composite primary keys.
A \deus{} account is identified by an OpenID \cite{ReRe2006}.
The optional application of XRIs\footnote{Extensible Resource Identifier,\\\url{http://www.oasis-open.org/committees/xri}} is in preparation.
Both identifier types can be used with XRDS\footnote{eXtensible Resource Descriptor Sequence,\\ \url{http://docs.oasis-open.org/xri/2.0/specs/}\\\url{xri-resolution-V2.0.html}} service discovery.
XRDS can be used for transfer protocol negotiation as it will later be elaborated in the description of the \deus{} transfer layer access.

The \deus{} storage and communication handles the artifacts in a general way, using the equivalent data artifacts of the \emph{generic layer}.
Any \dc{} is identified by a triple: The first component is a discriminator that is generated by the information provider.
The second component is the account identity of the information provider itself.
The third component is the account identity of the concerned person, about which the \dc{} provides information. 
%%
%The different IDs have diverse relevance among the phases ... For example, during repatriation, the concerned person's ID is used for addressing purposes and the information provider's ID is used for key resolution before signature verification.

A \emph{personal information file} (PIF) is identified only by the account identity of the concerned person it belongs to.
The \emph{foreign information file} (FIF) is identified by combining the concerned person's ID with the owning information consumer's ID.
Finally, the \emph{distributed information folder} (DIF) is owned and identified by the information consumer.

The actor-differentiated identification scheme is used by the persistence layer for each of the \deus{} artifacts.
It allows a \deus{} node to host multiple \deus{} accounts.
This provides multitenancy, which enables a proxy institution to professionally and securely host \deus{} accounts for multiple institutions.
In conclusion, \deus{} provides persistence that allows to scale discretionary from a central storage to a fully-distributed \deus{} node topology.

\subsection{From Cooperation to Collaboration}
\label{sect-closed-comm-env}

The basic \deus{} scenario as described above is primarily a cooperative one and not a collaborative one.
In cooperation, the participants are organizationally independent and the degree of interdependence remains low, although the information exchange is based on a mutual benefit.
In contrast, collaborative scenarios require a team to achieve collective results that the participants would be incapable of accomplishing when working alone.
Although collaborating partners remain organizationally independent, the degree of work-product interdependence is substantial.

Such a collaboration of healthcare professionals occurs for example in breast cancer treatment centers working with accredited partners.
Here, an enforced patient interaction would be counterproductive to the collaborative workflow and is not feasible.
Therefore, a specialized \deus{} scenario can be found:
Exchanging patient information without requiring a patient to decide about the distribution.
A patient account that exists only for logical purposes could be specially trimmed-down, so that it would auto-accept any repatriation requests from certified contributors.
The set of participants, comprising all accredited partners, could be templated and injected during application specific patient account creation as the list of contributers and subscribers.
For cancer treatment the oncologists, radiologists, and the post-operative care actually require \emph{all} available patient information unfiltered.
Thus, the phase in which the patient account performs subscriber selection can be switched off by configuring the according subsystem to publish any \dc{} to all subscribers.
Such a \emph{virtual patient account} should be hosted by the breast cancer treatment center.
%%
%Yet, the patient account should still be separated from the treatment center account because ... WHY?

An organizationally closed environment is mandatory for such virtual patient accounts.
Thus, this idea appears to be relevant for individual case environments, only, with moderate participant sets---not for life-long, large-scale patient files.
In life-long scenarios the patient must be taken into account as mediator.

%% file: 50_architecture.tex
%auto-ignore
\section{DEUS System Architecture}
\label{section-deus-architecture}

\begin{figure}[!bh] \centering 
\includegraphics[width=1\columnwidth]{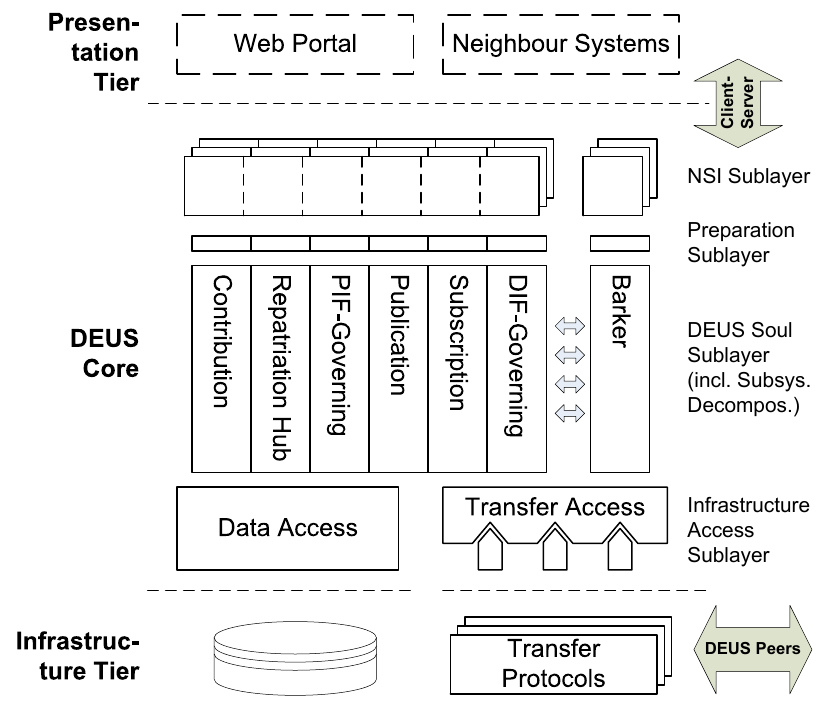}
    \caption{The DEUS architecture}
    \label{figure-DEUS-architecture}
\end{figure}

The \deus system is decomposed vertically into tiers and sublayers.
The horizontal decomposition into subsystems is deduced from the differentiation of \deus functional roles in Fig.\,\ref{figure-deus-roles}.
Therefore, the modules have high functional cohesion and a distinct scope of work.
As additional lateral decomposition, the Barker subsystem provides cross-cutting functionality for system-to-user interactions.
The resulting overall architecture is outlined in Fig.\,\ref{figure-DEUS-architecture}.

The neighbour systems in the presentation tier are not \deus{} peers but institutional systems that access their \deus{} system extension within an intranet by client/server access.
\deus peers communicate by arbitrary transfer protocols of the infrastructure tier.

\vspace{1cm}
\subsection{Component Model}

To enable clear interfaces and enforce module boundaries for side-effect free invocations, all pictured \uml{} components are realized as distinct \osgi\footnote{Open Services Gateway initiative, \url{http://www.osgi.org}} bundles.
The \osgi framework provides a runtime system to support a defined lifecycle, including hot deployment, installation, dependency resolution and service registration.
Component boundaries are controlled by explicitly specifying exported and imported interfaces and classes.
Apache Maven\footnote{\url{http://maven.apache.org}} was used as a build system to reflect the modularization in the project folder layout with Maven Multi-Modules and to support compile-time dependency resolution.

\subsection{Basic Decomposition}

The \emph{presentation tier} offers a web portal for user interaction.
Alternatively, neighbour systems inside the institutional intranet can access the \deus system by several remote invocation technologies that are provided by the \emph{neighbour systems interface} (NSI) sublayer.

%%% FORCING THE PICTURE AT THE SIDE OF SUBSECTION ``Soul Subsystems''
\begin{figure*}[ht] \centering 
\includegraphics[width=.85\textwidth]{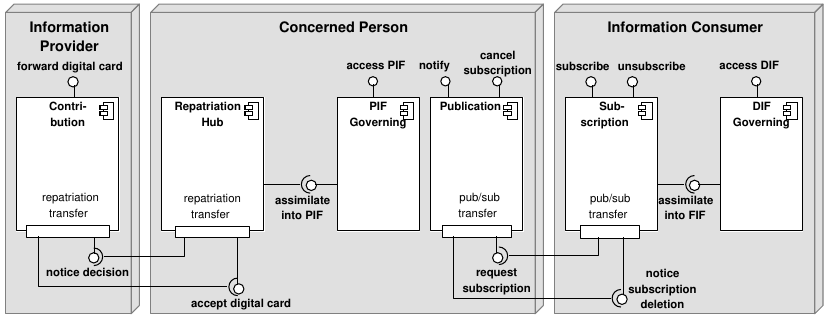}
    \caption{DEUS Soul subsystems (without the lateral Barker)}
    \label{figure-deus-subsystem-components}
\end{figure*}

The \emph{application tier} or \emph{core layer} provides the NSI sublayer as well as the preparation sublayer, the Soul and both infrastructure access sublayers.
The primarily disposed NSI remote invocation technology is a \rest{}\footnote{REpresentational State Transfer} \cite{Fiel2000} interface, since the \restful approach fits with the document-oriented perspective.
Modules that provide NSI access by \soap{} or \rmi{} technology could be deployed additionally.
All invocations to the subsystems that reside in the \deus \emph{Soul sublayer} are intercepted by the \emph{preparation sublayer}.
It checks for data validity and transforms the parameter types to domain data types, thereby shielding the Soul subsystems from calls with invalid data.
%%
%The Soul sublayer encompasses the core business functionality and is divided into subsystems that are elaborated later.

The \emph{infrastructure tier} encompasses data stores and \deus{} data schemas as well as transfer protocols for communication with remote \deus nodes.
The \emph{infrastructure access sublayer} within the \deus Core decouples from specific infrastructure technology.
The \emph{data access sublayer} provides an interface to the Soul that solely handles domain objects and is agnostic to any specific persistence technology.
An implementation module provides a binding to a dedicated back-end data store.
The \emph{transfer access sublayer} provides the Soul with a facility to communicate with remote \deus accounts.
It is transparent to the Soul, whether the targeted account resides on another or on the same \deus node.
%%
%The latter is possible, since the system provides multitenancy.

The next three sections will detail the notable parts of \deus{} architecture:
the Soul subsystems and their cooperation, the transfer access sublayer with its plug-in architecture for binding arbitrary message-based communication middleware, and the interaction mechanism between the end-user, the subsystems, and the messaging.

\subsection{Soul Subsystems}

Fig.\,\ref{figure-deus-subsystem-components} depicts the \deus{} Soul subsystem modules and their interfaces as well as their dependencies in cooperation.
For lucidity, the delegations to the lateral Barker system, that provides system-to-user interactions, are omitted.
The interplay between users, accounts, subsystems, and messages will be clarified later.
The subsystems are grouped by the higher-level distribution roles.
Each \deus{} node implements all roles.
The cooperation of the subsystem will be explained in the context of the con\-tri\-bu\-tion-re\-pa\-tri\-a\-tion-pub\-li\-ca\-tion phases.

The \emph{Contribution} subsystem provides a facility to contribute \dcs that have been exported by an HCIS from its \ehr into the \deus system.
%in form of a \texttt{\url{deus.core.soul.contribution:Contributer.forwardToCP(DigitalCard)}}
For this purpose a \texttt{forwardToCP(DigitalCard)} method is exported by the NSI to institutional neighbour systems.
The method takes a digital card, signs it, and forwards it to the concerned person's \deus{} account using the \deus{} peer transfer protocol infrastructure.
The account identity of the concerned person is part of the composite identity of the \dc that has to be repatriated.
The addressing is accomplished by using the account identity and resolving a transfer protocol ID by performing a handshake with the target account.
The communication itself will be detailed in the next section.
The \emph{Repatriation Hub} of the receiving account accepts the \dc{} by an \texttt{accept(DigitalCard)} method. 
This subsystem presents the decision, whether to accept or decline the contribution, to the user by delegating the event to the Barker subsystem which appends it to an attention list presented to the user.
The \dc is meanwhile persisted in a staging area.
If the user confirms the decision about the \dc acceptance, the Barker subsystem triggers the \emph{PIF-Governing} subsystem to migrate the \dc from the staging storage into the \pif storage (``\textsf{assimilate into PIF}'').
The PIF-Governing subsystem also exports its facility for \pif access to the user-interface or to institutional neighbour systems.

\begin{figure*}[ht] \centering 
\includegraphics[width=.6\textwidth]{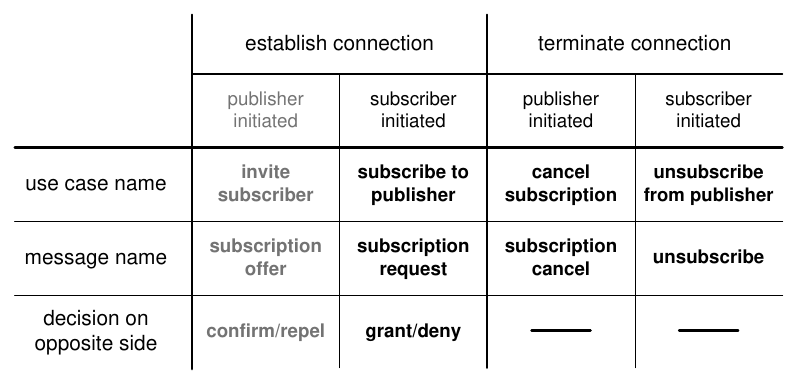}
    \caption{The DEUS vocabulary overview for pub/sub connection management}
    \label{figure-pub-sub-rel-terminology}
\end{figure*}

The functionality around the traditional pub/sub capabilities of the \deus system is comprised in the subsystems \emph{Publication} and \emph{Subscription} in Fig.\,\ref{figure-deus-subsystem-components}:
Establishing a publish-subscribe connection requires the subscriber to know the account ID of the publisher.
The Subscription module implements a \textsf{subscribe} interface that is exported by the NSI.
It results in a signed \textsf{request subscription} to the addressed publication subsystem.
The subscription request is delegated inside the Publication subsystem to the Barker, that presents the request to the user for decision.
The necessary command messages are mapped to the resolved transfer protocol.
The authentication relies on a pre-established PKI key exchange, which is discussed in the future work section.
At the moment, an establishment is based on the account IDs.
To complete the connection establishment at publication side, the publisher adds the subscriber ID to a \emph{list of accepted subscribers}.
The assignment of a subscriber into a specific publication group can be done during the acceptance of the request subscription. %; in any case, the subscriber is classified accordingly.
The decision is signed and sent back.
To complete the connection establishment at subscription side, the subscriber adds the publisher ID to its \emph{list of confirmed publishers}.
An overview over the \deus{} vocabulary for connection management is provided by fig.\,\ref{figure-pub-sub-rel-terminology}; the publisher initiated connection establishment is grey because it has not yet been implemented and is not represented by the current interfaces.

The last step in connection establishment concerns replication consistency:
The concerned person account initially publishes historic \dcs to the new consumer.
Different strategies can be applied:
First option would be the selection of \dcs that are always initially published to any accepted subscribers, containing for example the patient's master data and basic medical information like allergies or medication.
Second option would be the manual selection of \dcs by the concerned person, specifically for the new consumer.
Third option would be to remember a list of \dcs that have been published to a publication group, and to publish these \dcs to the new subscriber being added to the group. 
Another most trivial option would be to initially publish nothing.
The first option is use-case globally, group-independant, and instance-independant; the second is instance-specific; the third is group-specific; the forth is do-nothing.
More options, like sophisticated rule-based strategies, could be applied, but have not been evaluated at the moment.

Deleting an established publish-subscribe connection between two accounts can be initiated by the concerned person with a \textsf{cancel subscription} that is provided by the Publication subsystem.
It can also be initiated by the information consumer, with an \textsf{unsubscribe} that is provided by the Subscription subsystem.
The \deus system will delete the respective partner from the list of confirmed publishers or accepted subscribers.
The deletion of the \fif from the consumer actor's \dif will not be enforced; such decisions will be bound to judicial boundary conditions.
The concerned person can demand such deletion, either as part of the \textsf{cancel subscription} invocation or as part of the notification about the \textsf{unsubscribe}.
The Subscription subsystem, on the consumer side, will use its Barker to delegate the decision about the information deletion to its user.

The subsystem \emph{DIF-Governing} provides a facility to absorb a received digital card into the \fif.
Both, the DIF-Governing and PIF-Governing subsystems provide access to their governed data for the web portal or for institutional neighbour systems.

\subsection{Transfer Access}

\begin{figure}[ht] \centering 
\includegraphics[width=1\columnwidth]{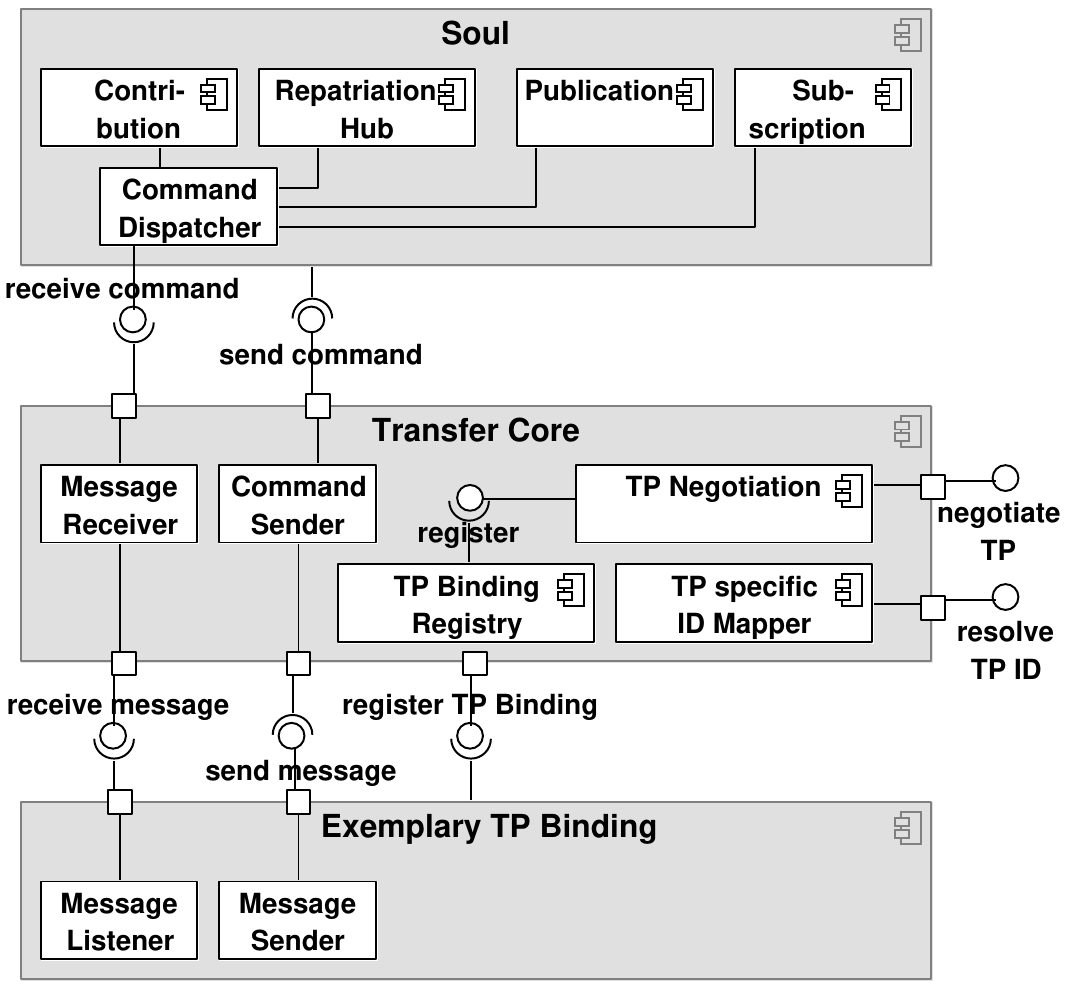}
    \caption{Inner architecture of the transfer access sublayer}
    \label{figure-deus-transfer}
\end{figure}

For being independent of a specific communication protocol, bindings to arbitrary protocols can be implemented to provide remote transfer.
This is reflected in the design of the transfer access subsystem which implements a plug-in architecture.
To decouple the Soul sublayer from a specific protocol binding the \emph{Transfer Core sublayer} is introduced.
This sublayer handles registration of transfer protocol bindings, negotiation of transfer protocols between two \deus{} nodes, and resolution of User IDs to transfer protocol IDs.
It is outlined in Fig.\,\ref{figure-deus-transfer}, together with an exemplary binding at the bottom.

\subsection{TP access layer mechanisms}

Available transfer protocol bindings are registered at the \textsf{TP Binding Registry} on startup and added to the list of available transfer protocols available for protocol negotiation.
If a subsystem from the Soul sublayer needs to transfer a message to another \deus account, it instruments the interface \textsf{send command} provided by Transfer Core.
The component \textsf{Command Sender} first negotiates the transfer protocol that should be used: 
The list of supported protocols is remotely retrieved from the receiver of the command.
These protocols are subsequently matched with the ones supported by the local \deus node and a common transfer protocol is chosen.
Both communication participants have priorities attached to their supported transfer protocols, which are taken into account during protocol negotiation.

Subsequently, the user ID of the remote \deus account is resolved to the transfer protocol ID of the negotiated transfer protocol using the remote offered interface \textsf{resolve TP ID}.
Following the \emph{Command Message} pattern in \cite[p.\,145]{HoWo2003}, the command to send and its parameters are marshalled and included in a message object.
The \textsf{send message} interface of the chosen transfer protocol binding is thereupon used to deliver the message.

Registered protocol bindings are listening for remote messages being sent by other \deus accounts.
If a message is received by a binding, binding makes a callback to the \textsf{receive message} interface of \textsf{Transfer Core}.
The message is unmarshalled, the contained command is extracted and passed to the Soul sublayer using its callback interface \textsf{receive command}.

The protocol bindings are realized as \osgi bundles, so that communication protocol bindings are registered at the startup of the \osgi bundle.
Using this approach, transfer protocols can be plugged in at runtime and thus are instantaneously available for selection during transfer protocol negotiation.

\subsection{Transfer Protocol Bindings}

%%% FORCING THE PICTURE AT THE SIDE OF SUBSECTION ``DEUS Communication Protocol''
\begin{figure*}[ht] \centering 
\includegraphics[width=0.85\textwidth]{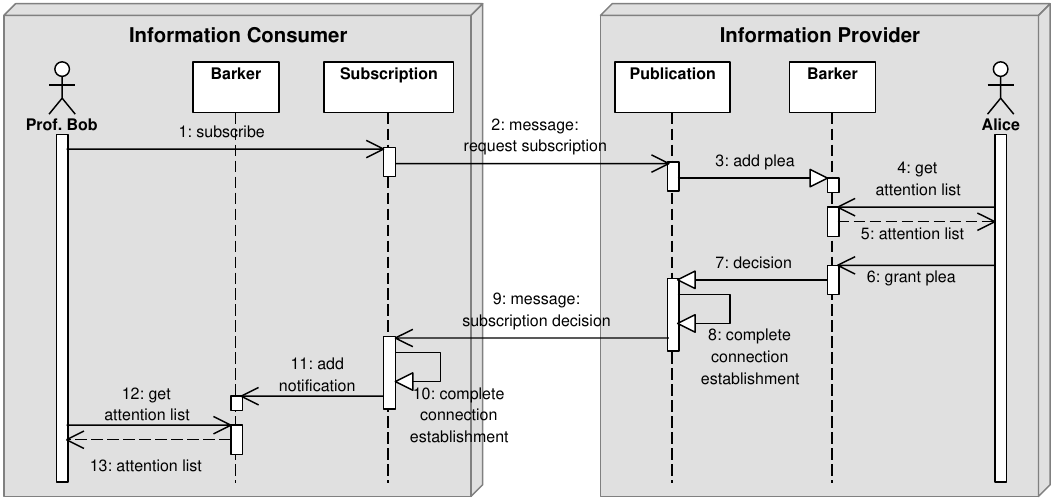}
    \caption{Request subscription sequence}
    \label{figure-deus-pub-sub-sequence}
\end{figure*}

The transfer layer for \deus{} account-to-account communication must support a \emph{Point-to-Point Channel} \cite[p.\,103]{HoWo2003} for repatriation and a \emph{Publish-Subscribe Channel} \cite[p.\,106]{HoWo2003}.
The most basic tranfer protocol binding is a local loopback protocol, for communication between user accounts that reside on the same \deus node.
The next binding module is based on a \rest{} delivery that implements a simple application layer multicast.
The \rest{}-based multicast requires a minimal \textsf{POST} interface that accepts messages.
Obviously, the simple \rest{} approach does not provide \emph{Guaranteed Delivery} \cite[p.\,122]{HoWo2003}.
Assuming that \deus nodes may fail and be offline for a certain period of time would require a sophisticated implementation of persistent queues.
Instead, another protocol binding is based on \xmpp{}\footnote{Extensible Messaging and Presence Protocol -- the core protocol of the Jabber instant messaging technology}:
The transfer access layer plug-in is based on an \xmpp{} client\footnote{\deus{} is using the Smack library} that acts as \emph{Messaging Mapper} \cite[p.\,477]{HoWo2003} and instruments a local \xmpp{} server\footnote{\deus is using OpenFire} as persistent queue.
The \xmpp{} server takes any measures that delay and guarantee the delivery of messages directed to these nodes during their downtime.

Further transfer protocol bindings are in consideration: JMS\footnote{Java Message Service, JSR-914} could be implemented with a JMS client instrumenting configurable JMS providers; messaging middleware like MQSeries/Websphere MQ or the Tibco service bus could augment the portfolio.
This is not supported by \rest and thus would need to be implemented using message queues on the publisher side.
Yet, the abstraction level of the protocol to be bound should ease the mapping of \deus{} high-level concepts:
\xmpp{} provides the abstraction of users and their accounts, being uniquely identified by an \xmpp{} ID, as well as concepts of subscription management, message routing, and delivery.
Support for a tranfer protocol inherent multicast is also provided -- by an \xmpp extension (XEP-0060) that extends the original  store-and-forward into a basic publish/subscribe message delivery.
Furthermore, \xmpp provides many official \xmpp Extension Protocols including support for service discovery, in-band account registration, search, message receipts, reliable data transfer, time-sensitive delivery, expiration of transient messages and more.

\subsection{User--Barker Interaction and Message Exchange}

The message protocol comprises command messages for establishment and termination of repatriation relationships, document messages for the repatriation as well as command messages for subscription management.
The document message for publication is the same as for the repatriation, the difference lies within the transferred \dc{} which is signed twice.
There are three types of command messages: request, decision, and notice messages.

Fig.\,\ref{figure-deus-pub-sub-sequence} outlines the interplay between the user actors, the Subscription/Publication subsystems, the messaging, and the Barker subsystem.
Prof.~Bob subscribes to Alice, invoking the \texttt{subscribe(UserID)} with Alice's account ID. % (e.g.~using a form in his local \deus{} web portal).
A \textsf{request subscription} message is sent to Alice's account and is dispatched to the Publication subsystem.
The transition from a command ($\rightarrow$ messaging) to an attention element ($\rightarrow$ human-interface) is done by the Publication subsystem that adds a \textsf{plea} object to the Barker.
The Barker integrates the plea as \emph{attention element} in the \emph{attention list}.
Alice, upon next login, displays the attention list and can grant or deny any plea.
The decision is delegated by the Barker to the Publication subsystem.
The example in Fig.\,\ref{figure-deus-pub-sub-sequence} describes the ``grant'' case, therefore the connection establishment is completed (in the ``deny'' case, any persistently staged objects for a request are cleaned up).
The decision is transfered to the requesting account, which performs its own request completion and adds the decision as \textsf{notification} to the Barker.
Again, the Barker absorbs the notification in the attention list.
Upon reading a notification, the user can mark it, so that it will not be displayed any more in the attention list
The Barker stores a complete history of any notification and decision elements.

%% file: 60_postarch.tex
%auto-ignore
\section{Distribution over Centralization}
\label{section-defense}

A risk in instrumenting a central content storage, like German D2D\footnote{Doctor to Doctor, \url{http://www.d2d.de}, based on PaDok cryptographic infrastructure \cite{PaDok}} or Google Health\footnote{\url{http://www.google.com/health}}, is an information leak that potentially involves all patients.
This is not comparable to any possible abuse scenario in today's paper-based infrastructure:
No current healthcare institution hoards information about so many patients as will do any centralized solution for inter-institutional scenarios.
The distributed approach mirrors the current state in paper-based working practice:
The patient information is available only to the directly involved healthcare systems.
As a result, the consequences of a security breach are limited to a fraction.
%
%%% any central solution would ... exhibit ... a blinking target for bad people, essential advantage of a distributed approach over 

Yet, even the smallest security breach still remains fatal due to the criticality of the involved information.
Hospital infrastructure commonly hosts electronic patient information and already applies profound security measures, 
e.g.\ \cite{VLC+2002}, but primary care participants might not be accustomed to the required security standards.
Therefore, a distributed approach has to allow proxy institutions to professionally and securely host accounts, i.e.\ for primary care participants or for patient participants.
For example, the health insurance funds or the associations of statutory health insurance physicians could provide account hosting. % in regional computing centers, providing access to their regional participants by virtual private networks (VPN). %, since such VPN infrastructure is today rolled out successively to primary care institutions by ... ToDo Rhön-Klinik Recherche / VPN bei uns in der Praxis? / Prof. Lenz \todocite{} ...
In conclusion, participating peer systems in the globally distributed healthcare environment are required to locally adhere to a multi-tenant data architecture as it is provided by \deus{}.

In medical care, the availability of information at the right time and at the right location (the ``point-of-care'') is crucial \cite{Ande1997}.
In contrast to the \deus{} push-based information distribution, a pull-based approach could allow a \hcis{} to query a patient account ad-hoc when information is needed.
Since a local replicate of the electronic patient file is absent, its advantages are lost:
These benefits include a reduced response time, because a remote call is avoided, which elevates end-user acceptance.
Furthermore, the absence of a local copy requires the continuous availability of the patient \deus{} node for information provision, which can not be guaranteed.
Nevertheless, even a pull-based approach requires a trust relationship to be built between the concerned person and the information consumer in advance, analogous to the \deus{} architecture.
In conclusion, the \deus{} push-based approach is more efficient than pull-based solutions.

\section{Related Work}
\label{section-related-work}

Besides data integration, the \emph{functional integration} has to provide operative interoperability between information systems.
Functional integration is addressed on a syntactic technical level and a semantic domain-specific level.
On the technical level, mature solutions exist, i.e.\,instrumenting component models like \ejb{} \cite{MeNe04testejb} and \dotnet{} or lightweight ones like \spring{} and \osgi{} as well as remote procedure calls like \iiop{}, \rmi{}, or \xml{}-based protocols like \soap{}.
Semantic functional integration requires interface and protocol standards for medical services.

Standards for electronic information exchange between the \emph{practice management systems} %(PMS) 
of the primary care and the hospitals and institutions of the secondary care are rare.
No universal exchange protocol and format exists for interchange of referral vouchers and discharge letters.
In Germany, the governmental project ``Elektronische Gesundheitskarte'' (cf.\,sect.\,\ref{sect-egk}) has not provided a solution for the issue since the project's outset in 2003.
Effective platforms like D2D require a central server for document handover.

Existing protocol standards for information exchange in distributed healthcare scenarios mainly focus on hospitals of the secondary care.
These standards encompass the earliest efforts in this area due to the complexity of a major hospital and its need for inter-sectional information exchange.
Available standards in healthcare, like \dicom{}\footnote{Digital Imaging and Communications in Medicine} \cite{BHPV1997}
or the \emph{cross-enterprise document sharing} (\xds{}) \cite{ITIv1} standard from \ihe{}\footnote{Integrating the Healthcare Enterprise, \url{www.ihe.net}} \cite{SiCh2001}, focus on the information exchange between hospital information systems (HIS) cooperating with ancillary systems like radiology information system (RIS), cardiology, and pathology systems, or laboratory information management system (LIMS).
Furthermore, there exist tailor-made regional integration efforts, but they are based on a central database system (DBS) with distributed transaction systems and diverse communication middleware.
Even wide-area RHIN architectures like \hygeianet{} \cite{TsKO2002} on the island Crete require a federated database schema \cite{KaTO2001} and are therefore tightly-coupled by their complex infrastructure being inadequate for transregional scaling.

Solving the information exchange in healthcare in a doc\-u\-ment-oriented fashion seems to be targeted by \ihe{} \xds{} which allows for distributed document repositories and access delegation.
For gaining experience with such standards, we implemented \emph{XdsRig}, an \ihe{} \xds{} test stand environment; it is published in \cite{NeWL09xdsrig} and available as open source.
Yet, in order to find documents in such a repository, a single central document registry is specified, reusing \ebxml{} registry methodology to provide a centralized method of indexing documents.
The central registry is a global system node that allows queries and that delegates the access to referenced documents to the original document repositories.
Such architecture targets complex hospitals with associated ancillary systems and is even applicable to regional integration efforts, but fails for nationwide application due to its centralized approach.

In \rhin{}s, several hospitals and ancillary institutions normally develop a shared set of IT services for information exchange.
Neutral organizations like \ihe{} try to establish interaction standards in format and protocol, little by little, based on standards like  \dicom{} or \hlseven{}.
Concurring standards currently in use can be placed into a classifying matrix of integration \cite{Lenz2006}.
In conclusion, a ``semantic gap'' is revealed, that is not covered by standards concerning the semantic integration.

Another term that describes \deus{} in regard to its patient-centered approach is \emph{Personal Health Record} (\phr{}), which is an old term that appeared first in 1978 \cite{britain1978computerisation}.
The term has gained new momentum.
However, solutions that are currently marketed as a \phr{} are Web-based and centralized; Sittig provides an overview in \cite{sittig2002phr}.
Well-known examples are systems like Dossia\footnote{Dossia is limited to employees of the few signee companies.}, PatientsLikeMe, Microsoft HealthVault\footnote{Microsoft HealthVault is not available in Germany.}, and the discontinued Google Health.
Whereas the original \phr{} concept naturally emphasizes on a paper perspective (like we do in \deus{}) the emerging Web-based \phr{} systems, however, are intrinsically database-oriented with semantically rich data models.
Rich data models imply integration efforts in a proprietary way and in a fragmented fashion.
In conclusion, they become ``yet another data silo'' or even a ``data tomb'':
When Google closed their Google Health service, the users got a database dump in form of an \xml{} export.
Granted, such an export was better than expected (and other \phr{} systems do not provide an export), but ultimately it is futile because other \phr{} vendors have incompatible data models and an import is not provided.
The \ihe{} started to attend issues of \phr{} integration by adopting it into their Patient Care Coordination (\textsc{pcc}) efforts \cite{ihe2010pcc}.
\deus{} can be considered as a distributed \phr{}; the document-orientation guarantees that it is possible, with the collection of electronic documents that constitute an \deus{} patient file, to export it into the file system at the user desktop at any time.

In the context of integrated ambient systems in healthcare there exist research projects for large-scale distributed middleware \cite{LSPS2008}.
They demonstrate that XMPP is even applicable to real-time healthcare scenarios with event-based body sensor networks.
Yet, the focus is completely different; they do not consider the three party scenario of \deus{} and the authoritative integration of the patient in information distribution.

\section{Future Work}

There are several open issues on different conceptual levels and technological layers.

\subsection{First Contact}

In \deus{}, a sophisticated solution for the first contact problem is required:
The OpenID identification of each \deus{} account could be integrated in a distributed master patient index (MPI) for the simplification of account connection establishment.
MPI systems like \ihe{} \pix{}\footnote{Patient Identifier Cross-Referencing} \cite{ITIv1} 
or \omg{}\footnote{Object Management Group} \pids{}\footnote{Patient IDentification Service} \cite{PIDS} 
instrument hierarchical federation with central system nodes and are not applicable in distributed environments.
Therefore, a loosely-coupled distributed patient identification service for inter-institutional purpose is required.

\subsection{Content Filtering}

An altogether open issue in \deus{} is \dc{} filtration because it requires a deep knowledge of medical taxonomies and ontologies.
An approach that enables subscribers to express predicates over the content of the \dc{}s could provide content-based subscriptions \cite{CaWo2002}.
Subscriber-defined predicates over well-known classification attributes of published \dc{}s would provide channel-based subscriptions. %, not to be misunderstood with the team-based subscriber-selection filter on publisher-side.
The same filter mechanisms could even be applied as further empowerment of the concerned person: 
Publisher-defined filters as a \emph{share set} for each individual subscriber could enable the concerned person with sophisticated filter mechanisms for extended control over published data.
Yet, any signed \dc{} can only be filtered completely or not at all; filtering information content of the repatriated \dc{} can only be applied for unsigned information with limited trust, for example contact information like telephone numbers.
However, diverse kinds of filtering \dc{} information would technically be possible, but would have tremendous unsolved medical and legal implications.
Singh et al. \cite{Sing08} are considering similar concepts for the English \ac{NHS}.

Another consideration about the content is the medical consistency of its information.
With the \textsc{oxdbs} project \cite{NeFL10oxdbs}, we extended a native \xml database system with validation by consistency checking of \textsf{OWL-DL} ontologies.
A possible integration of such a facility into our exchange platform seems promising.

\subsection{Group Management}

The subscriber selection requires some efforts from the concerned person, in form of physician group management.
Yet, there exists another approach to the subscriber selection: The contributor could provide reliable hints to the concerned person about which subscribers are appropriate in relation to a \dc{}.
Formalizing such hinting could automate the subscriber selection effectively; the PKI trust between the concerned person's and information provider's account has to be established for repatriation anyway.
% relying on the real-world trust between a patient in his or her physician in combination with the PKI trust which has to be established between the concerned person's and information provider's account for repatriation anyway.
%
In healthcare, the publication hints would require a classification system for the medical roles that a participant can take in treatment supply chains.

% \FIXME{The contribution and assimilation of a repatriated \dc{} into the \pif{} ... underlies a configurable strategy ... which decides about the merging semantics and the versioning of existing \dc{}s.}

\subsection{Process Model}

The document-oriented content distribution platform can be considered as a foundation for inter-institutional process support.
A distributed process model that adheres to the di\-ag\-nos\-tic-ther\-a\-peu\-tic cycle as coarse-grained intuitional reference from working practice is not yet available.
With a project named \dmps \cite{NeLe09dmps}, we worked on a platform that provides a shared process ID and maintains process history in order-entry and result-reporting scenarios.
The process history also provides information about the pretreatment or mutual treatment providers.
Additional characteristics for workflow support needs yet to be integrated into \deus{} like a formal model for pan-di\-ag\-nos\-tic and pan-ther\-a\-peu\-tic processes.

\subsection{Alignment to German e\textsc{gk}}
\label{sect-egk}

The German governmental project ``Elektronische Gesundheitskarte'' (e\textsc{gk}) was initiated after the German parliament passed a bill, the ``GKV-Modernisierungsgesetz'', to modernize health insurance cards in 2003.
The objective of the modernization plans was a telematics infrastructure for interconnecting healthcare facilities \cite{weichert2004egk}. With some new electronic health smartcards as its cryptographic foundation.

The primary e\textsc{gk} function is a public-key infrastructure (PKI) with smartcards and card readers as well as connectors for online communication.
On top of this basic communication platform, the \emph{mandatory e\textsc{gk} applications} comprise health insurance master data (e\textsc{gk} application \textsc{vsdd}: ``Versichertenstammdatendienst'') and electronic prescriptions (e\textsc{gk} application \textsc{vodd}: ``Verordnungsdatendienst'').
In addition, so called \emph{optional e\textsc{gk} applications} had been described in the official documentation \cite{egk2005fachlogisch}.
It comprised a requirement analysis for the following applications:
a patient's pharmaceuticals history (application \textsc{amdd}: ``Arzneimitteldokumentationsdienst''), 
the patient's emergency health data (application \textsc{nfdd}: ``Notfalldatendienst''), 
an infrastructure for physicians' result reporting (application \textsc{abd}: ``Arztbriefdienst''),
an infrastructure for physician-provided electronic patient files (application \textsc{epad}: ``Elektronischer Patientenaktendienst''), 
and an infrastructure for patient-provided information to his/her electronic file 
(application \textsc{pdd}: ``Patientendatendienst'').
No system specification actually exists to any of these optional e\textsc{gk} applications.
As mentioned in sect.\,\ref{section-related-work}, the e\textsc{gk} has not provided any solution, neither for its mandatory nor for its optional applications, since the project's outset in 2003.

In e\textsc{gk} terms, \deus{} represents a combined solution to \textsc{epad}\,\&\,\textsc{pdd}.
(Our co-project \dmps{} represents a solution to the \textsc{abd}.)
Yet, an alignment with the specific e\textsc{gk} concepts and technological infrastructure remains an open issue.

\subsection{Abstraction from Public-Key Infrastructures}

The research does not focus on PKI, but relies on existing PKIs, like e\textsc{gk}, in order to build authentication and authorization, during the establishment of contribute/repatriate or pub/sub relationships.
Yet, there exists competing PKI infrastructures in healthcare, like PaDok \cite{PaDok} or \ihe{} \atna{} \cite{ITIv1}.
In addition, an integration of general messaging middleware with role-based access control \cite{BEP+03} seems promising\footnote{For example, OASIS \cite{Baco02} is a role-based access control, with HERMES \cite{Piet02} as overarching project that integrates it.}, too.
To sign \dc{}s with arbitrary PKI technology, a generalized \deus{} PKI-component integration is required that provides independence of the various PKI infrastructures.

\subsection{Case Files}

Maintaining patient files requires a system extension at each participating site.
For ad-hoc establishment of data exchange without preceding system integration we have done considerable work on distributed case files.
The umbrella project is $\alpha$-Flow, where the central artifact is the $\alpha$-Doc that is a distributed case file.
The ``$\alpha$'' stands for ``active'' as in \emph{active documents} \cite{LED+1999}.
The general idea is to extend electronic documents with process status information to achieve workflow support.
Further information about the various aspects of the $\alpha$-Flow project can be found in \cite{NeLe09alphaflow,NeLe10alphaUC,ToNe11alphaprops,NSWL11alphaadaptive,WaNe12alphaoffsync,NeWL12offsync,NeHL12hydra,NeLe12alphaFlow}.
A seamless transition from $\alpha$-Flow case files to \deus{} patient files is prepared for by the conceptual similarity in their data model.
It would be required to include the considerations for case-related adaption of \deus{} (from sect.\,\ref{sect-closed-comm-env}) into an integrated patient-\&-case-file approach.
Hence, it remains an open issue to integrate both systems.

\section{Conclusion}

The \deus{} architecture achieves a document-oriented process support between strict autonomous institutions following the paper-based work practice as reference model.
For this purpose, \deus{} applies document-orientation based on a publish-subscribe design.
The essential argument for doc\-u\-ment-oriented integration over interface- or message-oriented integration lies in its capacity to support deferred system design.
Deferred system design is necessary for healthcare information systems due to their adaptive-evolutionary character.

The \deus{} architecture appoints the patient as integral participant of the information supply chain.
The warranty of data protection, imperative in healthcare, requires PKI integration on technological level but additionally requires a repatriation phase with end-user interception capability, resulting in a mediated publish-subscribe architecture.
The \deus{} architecture supports local autonomy, multi-tenant data architecture, platform independence, and loose coupling.

The current \deus{} platform has several unsolved aspects.
It requires concepts for the first contact problem, content filtering, and a connection to PKI infrastructure.
It would also benefit from a technical alignment with the German national e\textsc{gk} project.
\deus{} is a platform for patient files in which patients are empowered to control data distribution.
Support for individual cases requires that the patient file artefact model is refined one level, in the future, in order to model case files as subsets of a patient file.
Support for the planning and monitoring of case-centric treatment progress in form of work lists requires a document-oriented process model.
Furthermore, \deus{} is a light-weight system extension at each participating site and requires an installation setup.
In contrast, $\alpha$-Flow implements the active documents methaphor such that it can be deployed without prior software installation.
Currently, the $\alpha$-Flow approach to distributed case files, excluding the patient from the data flow, is better suited for true ad-hoc establishment of data exchange between healthcare professionals.
\deus{} is better suited for life-long patient files.
In conclusion, the current \deus{} implementation is a necessary first step towards a distributed platform for trans-sectional, long-term, patient-centered healthcare documentation.

%%
%The initial goal of the proposed \deus{} architecture is to foster the availability of patient information in order to bridge the gap between institutions of the primary and secondary care.
%%
%It supports inter-institutional teams and allows for patient-centered document distribution.
%%
%As subject of interchange \dc{}s maintain the self-contained character of a document, accentuating a finer information granularity than paper-based document practice.
%%
%In conclusion, \deus{} is a necessary first step towards a platform for distributed, trans-sectional, long-term, patient-centered healthcare documentation.

%% file: _FAU-CS-TR_deus_tex.bbl
\begin{thebibliography}{10}
\providecommand{\url}[1]{#1}
\csname url@samestyle\endcsname
\providecommand{\newblock}{\relax}
\providecommand{\bibinfo}[2]{#2}
\providecommand{\BIBentrySTDinterwordspacing}{\spaceskip=0pt\relax}
\providecommand{\BIBentryALTinterwordstretchfactor}{4}
\providecommand{\BIBentryALTinterwordspacing}{\spaceskip=\fontdimen2\font plus
\BIBentryALTinterwordstretchfactor\fontdimen3\font minus
  \fontdimen4\font\relax}
\providecommand{\BIBforeignlanguage}[2]{{%
\expandafter\ifx\csname l@#1\endcsname\relax
\typeout{** WARNING: IEEEtran.bst: No hyphenation pattern has been}%
\typeout{** loaded for the language `#1'. Using the pattern for}%
\typeout{** the default language instead.}%
\else
\language=\csname l@#1\endcsname
\fi
#2}}
\providecommand{\BIBdecl}{\relax}
\BIBdecl

\bibitem{leape1995saa}
L.~Leape, D.~Bates, D.~Cullen, J.~Cooper, H.~Demonaco, T.~Gallivan,
  R.~Hallisey, J.~Ives, N.~Laird, and G.~Laffel, ``{Systems analysis of adverse
  drug events. ADE Prevention Study Group},'' \emph{JAMA}, vol. 274, no.~1, pp.
  35--43, 1995.

\bibitem{Lenz2003a}
R.~Lenz and K.~Kuhn, ``Towards a continuous evolution and adaptation of
  information systems in healthcare,'' \emph{Int J Med Inf}, vol.~73, no.~1,
  pp. 75--89, July 2003.

\bibitem{Lenz2006}
R.~Lenz, M.~Beyer, and K.~Kuhn, ``Semantic integration in healthcare
  networks,'' \emph{Int J Med Inf}, vol.~76, no. 2-3, pp. 201--207, 2006.

\bibitem{NRDL09deus}
C.~P. Neumann, F.~Rampp, M.~Daum, and R.~Lenz, ``{A Mediated Publish-Subscribe
  System for Inter-Institutional Process Support in Healthcare},'' in
  \emph{Proc of the 3rd ACM Int'l Conf on Distributed Event-Based Systems (DEBS
  2009)}, Nashville, TN, USA, Jul. 2009.

\bibitem{WTB+2008}
C.~van Walraven, M.~Taljaard, C.~Bell, E.~Etchells, K.~Zarnke, I.~Stiell, and
  A.~Forster, ``{Information exchange among physicians caring for the same
  patient in the community},'' \emph{Canadian Medical Association Journal},
  vol. 179, no.~10, p. 1013, 2008.

\bibitem{LSLG200l}
K.~Lorig, D.~Sobel, D.~Laurent, and V.~Gonzalez, \emph{{Living a Healthy Life
  With Chronic Conditions: Self-management of Heart Disease, Arthritis,
  Diabetes, Asthma, Bronchitis, Emphysema \& Others}}.\hskip 1em plus 0.5em
  minus 0.4em\relax Bull Publishing Company, 2000.

\bibitem{TBBT2002}
M.~Tattersall, P.~Butow, J.~Brown, and J.~Thompson, ``{Improving doctors'
  letters.}'' \emph{Med J Aust}, vol. 177, no.~9, pp. 516--20, 2002.

\bibitem{Same2004}
\BIBentryALTinterwordspacing
J.~Samers, ``{Report on Integrated Care in Advanced Cancer Project},'' {Inner
  and Eastern Melbourne BreastCare Consortium}, Tech. Rep., mar 2004. [Online].
  Available:
  \url{http://www.health.vic.gov.au/breastcare/downloads/integratedcare.pdf}
\BIBentrySTDinterwordspacing

\bibitem{LBM+2005}
R.~Lenz, M.~Beyer, C.~Meiler, S.~Jablonski, and K.~Kuhn,
  ``{Informationsintegration in Gesundheitsversorgungsnetzen},''
  \emph{Informatik-Spektrum}, vol.~28, no.~2, pp. 105--119, 2005.

\bibitem{FMF+84}
R.~Fletcher, M.~O'Malley, S.~Fletcher, J.~Earp, and J.~Alexander,
  ``{{M}easuring the continuity and coordination of medical care in a system
  involving multiple providers},'' \emph{Med Care}, vol.~22, pp. 403--411, May
  1984.

\bibitem{MUBP2005}
M.~M{\"u}ller, F.~{\"U}ckert, T.~B{\"u}rkle, and H.~Prokosch,
  ``{Cross-institutional data exchange using the clinical document architecture
  (CDA)},'' \emph{International journal of medical informatics}, vol.~74, no.
  2-4, pp. 245--256, 2005.

\bibitem{WVVM2001}
M.~Williams, G.~Venters, and D.~Marwick, ``{Developing a regional healthcare
  information network},'' \emph{Information Technology in Biomedicine, IEEE
  Transactions on}, vol.~5, no.~2, pp. 177--180, 2001.

\bibitem{RoLa2003}
S.~Rothschild and S.~Lapidos, ``{Virtual Integrated Practice: Integrating Teams
  and Technology to Manage Chronic Disease in Primary Care},'' \emph{Journal of
  Medical Systems}, vol.~27, no.~1, pp. 85--93, 2003.

\bibitem{BKB+2003}
S.~Brucker, U.~Krainick, M.~Bamberg, B.~Aydeniz, U.~Wagner, A.~DuBois,
  C.~Claussen, R.~Kreienberg, and D.~Wallwiener, ``{Rationale, funktionelles
  Konzept, Definition und Zertifizierung},'' \emph{{Der Gyn{\"a}kologe}},
  vol.~10, p. 862, 2003.

\bibitem{PoBu2005}
J.~Powell and I.~Buchan, ``{Electronic Health Records Should Support Clinical
  Research},'' \emph{Journal of Medical Internet Research}, vol.~7, no.~1,
  2005.

\bibitem{schuler2002tutorial}
\BIBentryALTinterwordspacing
L.~Alschuler and K.~U. Heitmann, ``{CDA Introductory Tutorial},'' in \emph{HL7
  International CDA Conference}, Oct. 2002. [Online]. Available:
  \url{http://www.hl7.de/cda2002/absbiopres/cdaintro.pdf}
\BIBentrySTDinterwordspacing

\bibitem{sippel2005da}
B.~Sippel, ``{Evaluation und Integration von Standards zum Datenaustausch im
  medizinischen Umfeld},'' Diplomarbeit, Friedrich-Alexander-Universit{\"a}t
  Erlangen-N{\"u}rnberg, Mar. 2005.

\bibitem{MOD+1998}
C.~McDonald, J.~Marc~Overhage, P.~Dexter, B.~Takesue, and J.~Suico, ``{What is
  done, what is needed and what is realistic to expect from medical informatics
  standards},'' \emph{International Journal of Medical Informatics}, vol.~48,
  no. 1-3, pp. 5--12, 1998.

\bibitem{Zais1996}
A.~Zai{\ss}, \emph{{{\"U}berleitungstabelle zwischen ICD-9 und ICD-10}}.\hskip
  1em plus 0.5em minus 0.4em\relax Deutscher {\"A}rzte-Verlag, K{\"o}ln, 1996.

\bibitem{SPSW2001}
M.~Stearns, C.~Price, K.~Spackman, and A.~Wang, ``{SNOMED clinical terms:
  overview of the development process and project status.}'' in \emph{Proc AMIA
  Symp}, vol. 662, 2001, p.~6.

\bibitem{FMD+1996}
A.~Forrey, C.~McDonald, G.~DeMoor, S.~Huff, D.~Leavelle, D.~Leland, T.~Fiers,
  L.~Charles, B.~Griffin, F.~Stalling \emph{et~al.}, ``{Logical observation
  identifier names and codes (LOINC) database: a public use set of codes and
  names for electronic reporting of clinical laboratory test results},''
  \emph{Clinical Chemistry}, vol.~42, no.~1, pp. 81--90, 1996.

\bibitem{HL7v26}
{HL7v2}, ``{ANSI/HL7 V2.6-2007},''
  \url{http://www.hl7.org/Library/standards_non1.htm}.

\bibitem{DAB+2001}
R.~Dolin, L.~Alschuler, C.~Beebe, P.~Biron, S.~Boyer, D.~Essin, E.~Kimber,
  T.~Lincoln, and J.~Mattison, ``{The HL7 Clinical Document Architecture},''
  \emph{Journal of the American Medical Informatics Association}, vol.~8,
  no.~6, pp. 552--569, 2001.

\bibitem{HeSD2003}
K.~Heitmann, R.~Schweiger, and J.~Dudeck, ``{Discharge and referral data
  exchange using global standards -- the SCIPHOX project in Germany},''
  \emph{International Journal of Medical Informatics}, vol.~70, no. 2-3, pp.
  195--203, 2003.

\bibitem{eArztbrief}
{eArztbrief -- D2D Telematik-Plattform der Kassen{\"a}rztlichen Vereinigungen},
  ``{Elektronischer Arztbrief im D2D-System},''
  \url{http://www.d2d.de/index.php?id=17}, 2005.

\bibitem{LeKu2003}
R.~Lenz and K.~Kuhn, ``{A strategic approach for business-IT alignment in
  health information systems},'' \emph{Lecture notes in computer science}, pp.
  178--195, 2003.

\bibitem{Pate2002}
N.~Patel, \emph{{Adaptive Evolutionary Information Systems}}.\hskip 1em plus
  0.5em minus 0.4em\relax Idea Group Inc, 2002.

\bibitem{HoWo2003}
G.~Hohpe and B.~Woolf, \emph{{Enterprise integration patterns: Designing,
  building, and deploying messaging solutions}}.\hskip 1em plus 0.5em minus
  0.4em\relax Addison-Wesley Longman Publishing Co., Inc. Boston, MA, USA,
  2003.

\bibitem{RoSS2007}
\BIBentryALTinterwordspacing
S.~Rozsnyai, J.~Schiefer, and A.~Schatten, ``{Concepts and Models for Typing
  Events for Event-Based Systems},'' in \emph{International conference on
  Distributed event-based systems (DEBS'07)}, 2007. [Online]. Available:
  \url{http://cocoon.ifs.tuwien.ac.at/pub/debs/debs2007.pdf}
\BIBentrySTDinterwordspacing

\bibitem{Lenz2009yb}
R.~Lenz, ``{Information Systems in Healthcare -- state and steps towards
  sustainability},'' in \emph{IMIA Yearbook of Medical Informatics},
  A.~Geissbuhler and C.~Kulikowski, Eds.\hskip 1em plus 0.5em minus 0.4em\relax
  Stuttgart: Schattauer, 2009, accepted for publication.

\bibitem{HigginsProject}
{Eclipse project}, ``{Higgins Open Source Identity Framework},''
  \url{http://www.eclipse.org/higgins/}.

\bibitem{euministererklaerung2003}
eHealth, ``{Ministerial Declaration, Brussels},''
  \url{http://europa.eu.int/information_society/eeurope/ehealth/conference/2003/doc/min_dec_22_may_03.pdf},
  May 2003.

\bibitem{CaWo2002}
A.~Carzaniga and A.~Wolf, ``{Content-Based Networking: A New Communication
  Infrastructure},'' \emph{Lecture Notes in Computer Science}, vol. 2538, pp.
  59--68, 2002.

\bibitem{ReRe2006}
D.~Recordon and D.~Reed, ``{OpenID 2.0: a platform for user-centric identity
  management},'' in \emph{{Proceedings of the second ACM workshop on Digital
  identity management}}.\hskip 1em plus 0.5em minus 0.4em\relax ACM Press New
  York, NY, USA, 2006, pp. 11--16.

\bibitem{Fiel2000}
R.~Fielding, ``{Architectural Styles and the Design of Network-based Software
  Architectures},'' Ph.D. dissertation, University of California, 2000.

\bibitem{PaDok}
{Fraunhofer Institut f{\"u}r Biomedizinische Technik (IBMT)}, ``{PaDok --
  Patientenbegleitende Dokumentation},''
  \url{http://www.ibmt.fraunhofer.de/fhg/Images/MT_padoknetzkonzept_de_tcm266-68980.pdf},
  2000.

\bibitem{VLC+2002}
J.~Vazquez-Naya, J.~Loureiro, J.~Calle, J.~Vidal, and A.~Sierra, ``{Necessary
  security mechanisms in a PACS DICOM access system with Web technology},''
  \emph{Journal of Digital Imaging}, vol.~15, pp. 107--111, 2002.

\bibitem{Ande1997}
J.~Anderson, ``{Clearing the way for physicians' use of clinical information
  systems},'' \emph{Communications of the ACM}, vol.~40, no.~8, pp. 83--90,
  1997.

\bibitem{MeNe04testejb}
M.~Meyerh{\"o}fer and C.~Neumann, ``{TestEJB} -- a measurement framework for
  {EJBs},'' in \emph{Proc of the 7th Int'l Symposium on Component-Based
  Software Engineering (CBSE'04) in conjunction with the 26th Int'l Conf on
  Software Engineering (ICSE'04)}, ser. Lecture Notes in Computer Science,
  I.~Crnkovic, Ed., vol. 3054.\hskip 1em plus 0.5em minus 0.4em\relax
  Edinburgh, UK: Springer, Berlin, DE, May 2004, pp. 294--301.

\bibitem{BHPV1997}
W.~Bidgood, S.~Horii, F.~Prior, and D.~Van~Syckle, ``{Understanding and Using
  DICOM, the Data Interchange Standard for Biomedical Imaging},'' \emph{Journal
  of the American Medical Informatics Association}, vol.~4, no.~3, pp.
  199--212, 1997.

\bibitem{ITIv1}
\BIBentryALTinterwordspacing
{ACC}, {HIMSS}, and {RSNA}, \emph{{IHE IT Infrastructure Technical Framework,
  vol. 1 (ITI TF-1): Integration Profiles, Rev 4.0}}, IHE Std., Rev. 4.0, Aug.
  2007. [Online]. Available:
  \url{http://www.ihe.net/Technical_Framework/upload/IHE_ITI_TF_4_0_Vol1_FT_2007_08_22.pdf}
\BIBentrySTDinterwordspacing

\bibitem{SiCh2001}
E.~Siegel and D.~Channin, ``{Integrating the Healthcare Enterprise: A Primer 1
  Part 1. Introduction},'' \emph{RadioGraphics}, vol.~21, no.~5, pp.
  1339--1341, 2001.

\bibitem{TsKO2002}
M.~Tsiknakis, D.~Katehakis, and S.~Orphanoudakis, ``{An open, component-based
  information infrastructure for integrated health information networks},''
  \emph{International Journal of Medical Informatics}, vol.~68, no. 1-3, pp.
  3--26, 2002.

\bibitem{KaTO2001}
D.~Katehakis, M.~Tsiknakis, and S.~Orphanoudakis, ``{Enabling Components of
  HYGEIAnet},'' in \emph{Proc. of TEPR}, 2001, pp. 146--153.

\bibitem{NeWL09xdsrig}
C.~P. Neumann, F.~Wagner, and R.~Lenz, ``{XdsRig -- Eine Open-Source IHE XDS
  Testumgebung},'' in \emph{Tagungsband der 54.\,GMDS-Jahrestagung}.\hskip 1em
  plus 0.5em minus 0.4em\relax Essen, DE: {Deutsche Gesell\-schaft f{\"u}r
  Medizinische Informatik, Biometrie und Epidemiologie} (GMDS), Sep. 2009.

\bibitem{britain1978computerisation}
G.~Britain, ``Computerisation of personal health records,'' \emph{Health
  Visit}, vol.~51, p. 227, 1978.

\bibitem{sittig2002phr}
D.~F. Sittig, ``Personal health records on the internet: a snapshot of the
  pioneers at the end of the 20th century,'' \emph{International Journal of
  Medical Informatics}, vol.~65, no.~1, pp. 1--6, 2002.

\bibitem{ihe2010pcc}
{IHE: Integrating the Healthcare Enterprise}, ``{IHE Patient Care Coordination
  (PCC), Technical Framework, Volume 1, Revision 6.0},''
  \url{www.ihe.net/Technical_framework/upload/IHE_PCC_TF_Rev6-0_Vol_1_2010-08-30.pdf},
  Aug. 2010.

\bibitem{LSPS2008}
W.~Labidi, J.~Susini, P.~Paradinas, and M.~Setton, ``{XMPP based Health Care
  Integrated Ambient Systems Middleware},'' in \emph{Developing Ambient
  Intelligence: Proceedings of the Second International Conference on Ambient
  Intelligence Developments (Ami. D'07)}.\hskip 1em plus 0.5em minus
  0.4em\relax Springer, 2008, p.~92.

\bibitem{PIDS}
{Object Management Group, Inc. (OMG)}, ``{Person Identification Service (PIDS)
  Specification},''
  \url{http://www.omg.org/technology/documents/formal/index.htm}, April 2001.

\bibitem{Sing08}
J.~Singh, L.~Vargas, J.~Bacon, and K.~Moody, ``{Policy-Based Information
  Sharing in Publish/Subscribe Middleware},'' in \emph{IEEE Workshop on
  Policies for Distributed Systems and Networks (POLICY)}, 2008, pp. 137--144.

\bibitem{NeFL10oxdbs}
C.~P. Neumann, T.~Fischer, and R.~Lenz, ``{OXDBS -- Extension of a native XML
  Database System with Validation by Consistency Checking of OWL-DL
  Ontologies},'' in \emph{Proc of the 14th International Database Engineering
  \& Applications Symposium (IDEAS'10)}, Montreal, QC, CA, Aug. 2010.

\bibitem{NeLe09dmps}
C.~P. Neumann and R.~Lenz, ``{A Light-Weight System Extension Supporting
  Document-based Processes in Healthcare},'' in \emph{Proc of the 3rd Int'l
  Workshop on Process-oriented Information Systems in Healthcare (ProHealth'09)
  in conjunction with the 7th Int'l Conf on Business Process Management
  (BPM'09)}, Ulm, DE, Sep. 2009.

\bibitem{weichert2004egk}
T.~Weichert, ``{Die elektronische Gesundheitskarte},'' \emph{Datenschutz und
  Datensicherheit}, vol.~28, no.~7, pp. 391--403, 2004.

\bibitem{egk2005fachlogisch}
\BIBentryALTinterwordspacing
{Fraunhofer Institut}, ``{Fachlogische Modellierung und spezifische
  Anwendungsdienste der elektronischen Gesundheitskarte},'' Mar. 2005.
  [Online]. Available:
  \url{http://www.ehealthopen.com/FuE/PDF/eGK-Fachmodlanw-v1.0.pdf}
\BIBentrySTDinterwordspacing

\bibitem{BEP+03}
A.~Belokosztolszki, D.~Eyers, P.~Pietzuch, J.~Bacon, and K.~Moody,
  ``{Role-Based Access Control for Publish/Subscribe Middleware
  Architectures},'' in \emph{2nd International Workshop on Distributed
  Event-Based Systems (DEBS'03)}, 2003, pp. 1--8.

\bibitem{Baco02}
J.~Bacon, K.~Moody, and W.~Yao, ``{A model of OASIS role-based access control
  and its support for active security},'' \emph{ACM Transactions on Information
  and System Security (TISSEC)}, vol.~5, no.~4, pp. 492--540, 2002.

\bibitem{Piet02}
P.~Pietzuch and J.~Bacon, ``{Hermes: A Distributed Event-Based Middleware
  Architecture},'' in \emph{1st International Workshop on Distributed
  Event-Based Systems (DEBS)}, 2002.

\bibitem{LED+1999}
A.~LaMarca, W.~Edwards, P.~Dourish, J.~Lamping, I.~Smith, and J.~Thornton,
  ``{Taking the work out of workflow: mechanisms for document-centered
  collaboration},'' in \emph{Proceedings of the sixth conference on European
  Conference on Computer Supported Cooperative Work}.\hskip 1em plus 0.5em
  minus 0.4em\relax Kluwer Academic Publishers Norwell, MA, USA, 1999, pp.
  1--20.

\bibitem{NeLe09alphaflow}
C.~P. Neumann and R.~Lenz, ``{alpha-Flow: A Document-based Approach to
  Inter-Institutional Process Support in Healthcare},'' in \emph{Proc of the
  3rd Int'l Workshop on Process-oriented Information Systems in Healthcare
  (ProHealth'09) in conjunction with the 7th Int'l Conf on Business Process
  Management (BPM'09)}, Ulm, DE, Sep. 2009.

\bibitem{NeLe10alphaUC}
------, ``{The alpha-Flow Use-Case of Breast Cancer Treatment -- Modeling
  Inter-Institutional Healthcare Workflows by Active Documents},'' in
  \emph{Proc of the 8th Int'l Workshop on Agent-based Computing for Enterprise
  Collaboration (ACEC) at the 19th Int'l Workshops on Enabling Technologies:
  Infrastructures for Collaborative Enterprises (WETICE 2010)}, Larissa, GR,
  Jun. 2010.

\bibitem{ToNe11alphaprops}
A.~Todorova and C.~P. Neumann, ``{alpha-Props: A Rule-Based Approach to `Active
  Properties' for Document-Oriented Process Support in Inter-Institutional
  Environments},'' in \emph{Lecture Notes in Informatics (LNI) Seminars 10 /
  Informatiktage 2011}, L.~Porada, Ed.\hskip 1em plus 0.5em minus 0.4em\relax
  Gesellschaft f{\"u}r Informatik e.V. (GI), Mar. 2011.

\bibitem{NSWL11alphaadaptive}
C.~P. Neumann, P.~K. Schwab, A.~M. Wahl, and R.~Lenz, ``{alpha-Adaptive:
  Evolutionary Workflow Metadata in Distributed Document-Oriented Process
  Management},'' in \emph{Proc of the 4th Int'l Workshop on Process-oriented
  Information Systems in Healthcare (ProHealth'11) in conjunction with the 9th
  Int'l Conf on Business Process Management (BPM'11)}, Clermont-Ferrand, FR,
  Aug. 2011.

\bibitem{WaNe12alphaoffsync}
A.~M. Wahl and C.~P. Neumann, ``{alpha-OffSync: An Offline-Capable
  Synchronization Approach for Distributed Document-Oriented Process Management
  in Healthcare},'' in \emph{Lecture Notes in Informatics (LNI) Seminars 11 /
  Informatiktage 2012}, L.~Porada, Ed.\hskip 1em plus 0.5em minus 0.4em\relax
  Gesellschaft f{\"u}r Informatik e.V. (GI), Mar. 2012.

\bibitem{NeWL12offsync}
C.~P. Neumann, A.~M. Wahl, and R.~Lenz, ``{Adaptive Version Clocks and the
  OffSync Protocol},'' in \emph{Proc of the 10th IEEE Int'l Symposium on
  Parallel and Distributed Processing with Applications (ISPA-12)}, Madrid,
  Spain, Jul. 2012, accepted for publication.

\bibitem{NeHL12hydra}
C.~P. Neumann, S.~A. Hady, and R.~Lenz, ``{Hydra Version Control System},'' in
  \emph{Proc of the 10th IEEE Int'l Symposium on Parallel and Distributed
  Processing with Applications (ISPA-12)}, Madrid, Spain, Jul. 2012, accepted
  for publication.

\bibitem{NeLe12alphaFlow}
C.~P. Neumann and R.~Lenz, ``{The alpha-Flow Approach to Inter-Institutional
  Process Support in Healthcare},'' \emph{International Journal of
  Knowledge-Based Organizations (IJKBO)}, vol.~2, no.~3, 2012, accepted for
  publication.

\end{thebibliography}
